\documentclass[preprint,pre,showkeys,showpacs]{revtex4}
\newcommand{\beq}{\begin{equation}}
\newcommand{\eeq}{\end{equation}}
\usepackage{graphicx}
\usepackage{graphics}
\usepackage{latexsym}
\usepackage{subfigure}
\usepackage{multirow}
\usepackage{dcolumn}
\begin{document}

\title{Spatiotemporal intermittency and scaling laws in the coupled sine
circle map lattice}

\author{Zahera Jabeen}
\email{zahera@physics.iitm.ac.in}
\author{Neelima Gupte}
\email{gupte@physics.iitm.ac.in}
\affiliation{Indian Institute of Technology-Madras, Chennai, India}
\keywords{Spatiotemporal intermittency, Scaling exponents, Directed Percolation, Spatial Intermittency}
\date{\today}
\pacs{05.45.Ra, 05.45.-a, 05.45.Df, 64.60.Ak}
\begin{abstract}

We study spatio-temporal intermittency (STI) in a system of coupled sine
circle maps. The phase diagram of the system shows parameter regimes
with STI of both the directed percolation (DP) and non-DP class. STI
with synchronized laminar behaviour belongs to the DP class. The regimes
of  non-DP behaviour show spatial intermittency (SI), where the
temporal behaviour of both the laminar and burst regions is regular, and the
distribution of laminar lengths scales as a power law. The regular temporal
behaviour for the bursts seen in these regimes of spatial intermittency can be
periodic or quasi-periodic, but the laminar length distributions scale
with the same power-law, which is distinct from the DP case. STI with traveling wave (TW) laminar states also appears in the phase diagram. Soliton-like structures appear in this regime. These are responsible for cross-overs with accompanying non-universal exponents. The
soliton lifetime distributions show power law scaling in regimes of long average
soliton life-times, but  peak at characteristic scales with a power-law tail in
regimes of short average soliton life-times. The signatures of each type of intermittent behaviour can be found in the dynamical characterisers of the system viz. the
eigenvalues of the stability matrix. We discuss the implications of our results for behaviour seen in other systems which exhibit spatio-temporal intermittency.

\end{abstract}
\maketitle

\section{Introduction}
The phenomena of spatiotemporal intermittency (STI), wherein laminar
states which exhibit regular temporal behaviour  co-exist in space and
time with burst states of irregular dynamics, is ubiquitous in both
natural and experimental systems. Such behaviour has been seen in
experiments on convection \cite{cili,daviaud}, counterrotating
Taylor-Couette flows \cite{colo},  oscillating ferro-fluidic spikes
\cite{rupp}, experimental and numerical  studies of rheological fluids
\cite{sriram, fielding}, and in experiments on hydrodynamic columns
\cite{pirat}. In theoretical studies, STI has been seen in PDEs such as
the damped Kuramoto-Sivashinsky equation \cite{kschate} and the
one-dimensional Ginzburg Landau equation \cite{glchate}, in coupled map
lattices \cite{kaneko} such as the Chat\'e-Manneville CML \cite{Chate},
the inhomogeneously coupled logistic map lattice \cite{ash}, and in
cellular automata studies.   

A variety of scaling laws have been observed in these systems. However,
there are no definite conclusions about their universal behaviour. The
type of spatiotemporal intermittency in which a laminar site becomes
active (turbulent) only if at least one of its neighbour is active, has
been conjectured to belong to the directed percolation (DP) universality
class \cite{Pomeau}. The dry state or the absorbing state in DP is
identified with the laminar state in STI, and the wet state of DP
corresponds  to the active state in STI, with time as the directed axis.
However, a CML specially designed to exhibit STI by Chat\'e and
Manneville showed critical exponents significantly different  from the
DP universality class \cite{Chate}. This led to a long debate in the
literature \cite{grassberger,bohr,rolf,jan}.                   
It was concluded that the presence of coherent
structures, called solitons, were responsible for spoiling the 
analogy with DP. 
The nature of the transition to STI and the identification of the
universality classes of STI is still an
unresolved issue, and is a topic of current interest.

Earlier studies of the diffusively coupled sine circle map lattice showed  regimes of
STI  which were completely free of solitons \cite{jan,
zjngpre}. Two types of STI were seen along the bifurcation boundaries of
the bifurcation from the synchronized solution. The first type of STI
showed an  entire set of static and dynamic scaling exponents which matched
with the DP exponents, and therefore was seen to belong convincingly to
the DP class. The other type of intermittency, 
where                                   
both the laminar and burst regions showed                        
regular temporal dynamics, was called spatial intermittency (SI). The laminar length distribution for this case showed 
characteristic power-law behaviour with its own characteristic exponent $\zeta=1.1$. This kind of behaviour
has been observed in the sine circle    
map lattice as well as in the inhomogenously coupled logistic map
lattice. In the case of the sine circle map lattice, both types of
intermittency, viz. STI of the DP 
class, and SI which does not belong to the DP class, were seen in 
different regions of the  phase diagram. Moreover, distinct signatures of the 
two types of behaviour were picked up by the dynamical characterisers of the
system, i.e. the eigenvalues of the stability matrix. The eigenvalue
spectrum was continuous in the DP regime, but exhibited the presence of gaps
in the SI regime.  

Different types of behaviour are seen within the SI class itself. 
The laminar state is synchronized in nature, but the burst state can
be periodic or quasi-periodic in its dynamical behaviour.
The periodic burst states can have different temporal periods. Burst states of the traveling wave type are observed at several points in the phase diagram.
The distribution of laminar lengths shows power-law scaling in both
cases with the same exponent.

The SI regimes lie close to the bifurcation boundaries of the synchronized 
solutions. The SI with traveling wave (TW) bursts bifurcates further via 
tangent-period doubling bifurcations, to STI with TW laminar states and 
turbulent bursts. This type of STI is contaminated with coherent structures 
similar to the solitons that spoil the DP regime in the Chat\'e-Manneville CML.
The solitons induce cross-over behaviour and the exponents in this regime take
non-universal values. The distribution of soliton lifetimes shows two 
characteristic regimes. In the short soliton lifetime regime, the distribution 
shows a peak which indicates the presence of a characteristic time scale, and 
has a power-law tail. In the longer soliton lifetime regime, the distribution 
has no characteristic scale and shows pure power-law behaviour. The solitons in this
regime also change the order of the phase transition in the system. 
 
The dynamical characterisers of the
system show signatures of the different types of temporal behaviour of
the burst states.
As mentioned earlier, the eigen-value distribution of the stability 
matrix is gapless for the STI with DP exponents, whereas
distinct gaps are seen in the distribution 
for the SI class. The number of gaps in the eigenvalue distribution of SI
as a function of bin size shows power-law behaviour. However, the
scaling exponent is different for SI with quasi-periodic bursts and SI
with periodic bursts.
We discuss the 
implications of our results for behaviour seen in
other systems which exhibit spatiotemporal intermittency.

The organisation of this paper is as follows. Section II gives the details of the model and the phase diagram obtained. The two universality classes of spatiotemporal intermittency seen in this system, as well as the variations within the SI class are discussed in section III. Section IV explains the role played by the solitons in inducing a cross-over behaviour in STI with traveling wave laminar state. The signatures of each type of intermittent behaviour is seen in the dynamical characterisers of the system. This has been discussed in section V. We conclude
with a discussion of these results and their implications for other systems.

\section{The model and the phase diagram}
\begin{figure}[!t]
\vspace{-.1in}
\begin{center}
\includegraphics[height=2.7in,width=3.8in]{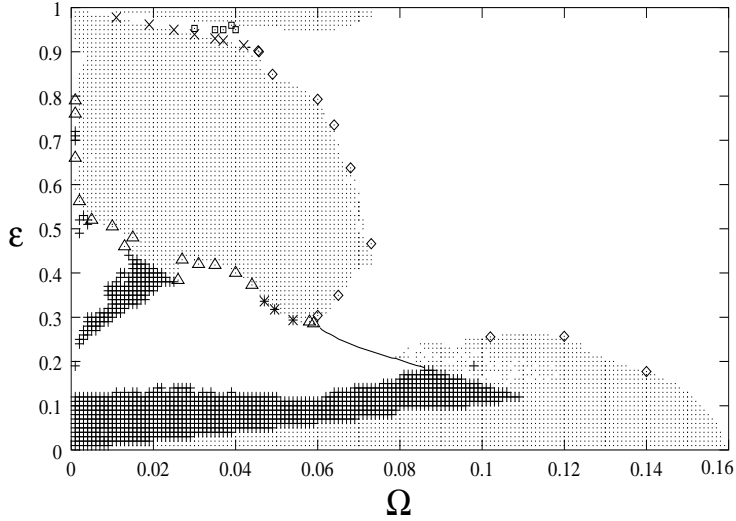}
\vspace{-.1in}
\vspace{-.2in}

\end{center}
\caption{ shows the phase diagram for the coupled sine circle map lattice evolved using random initial conditions. The spatiotemporally synchronized solutions are represented by dots. The points at which DP exponents have been obtained are marked by diamonds ($\Diamond$). At points marked with triangles ($\triangle$), SI with quasiperiodic bursts is seen. SI with TW bursts is seen at points marked by crosses ($\times$) and SI with period-5 bursts are seen at points marked with asterisks ($\ast$). STI with TW laminar states and solitons is seen at the points marked by boxes ($\Box$). The cluster solutions are marked with plus (+) signs \label{pd}.}
\end{figure}

The coupled sine circle map lattice is defined by the evolution equations
\beq
x_i^{t+1}=(1-\epsilon)f(x_i^t)+\frac{\epsilon}{2}[ f(x_{i-1}^t) + f(x_{i+1}^t) ]
\pmod{1}
\label{evol}
\eeq
where $i=1,\ldots,N$ and $t$ are the discrete site and time indices respectively
with $N$ being the size of the system,
and $\epsilon$ being the strength of the coupling between the site $i$ and its two nearest neighbours. The local on-site map, $f(x)$ is the sine circle map defined as 
\beq
f(x)=x+\Omega-\frac{K}{2\pi}\sin(2\pi x)
\label{sine}
\eeq
Here, $K$ is the strength of the nonlinearity and $\Omega$ is the
winding number of the  single sine circle map in the absence of the
nonlinearity. The coupled sine circle map lattice has been known to
model the mode-locking behaviour \cite{gauri2} seen commonly in coupled
oscillators, Josephson Junction arrays, etc, and is also found to be
amenable to analytical studies \cite{Nandini}. The phase diagram of this
system is highly sensitive to initial conditions due to the presence of
many degrees of freedom.  Studies of this model for several classes of
initial conditions have yielded rich phase diagrams with many distinct types of attractors \cite{Nandini, gauri2}. 

We study the system with random initial conditions. The system is updated synchronously with periodic boundary conditions in the parameter regime  $0 < \Omega < \frac{1}{2\pi}$ and $K=1$ (where the single circle map has temporal period 1 solutions in this regime); the coupling strength, $\epsilon$ is varied from $0$ to $1$. 

\begin{figure}
\begin{center}
\vspace{-.6in}
\begin{tabular}{cc}
(a)&(b)\\
\vspace{-1.3cm} 
{\hspace{-.5in}\includegraphics[scale=.8]{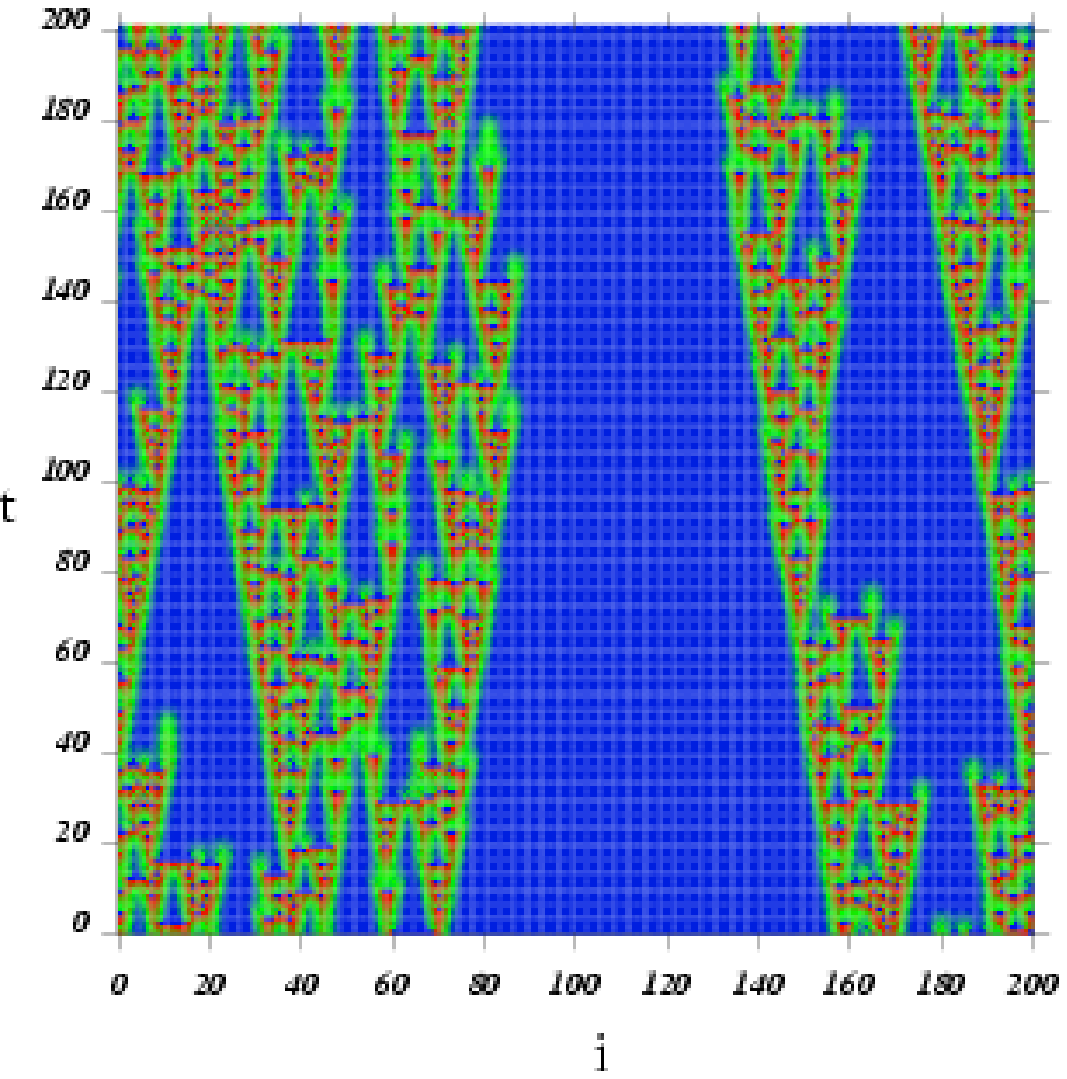}
}&
{\hspace{-.5cm}\includegraphics[scale=.8]{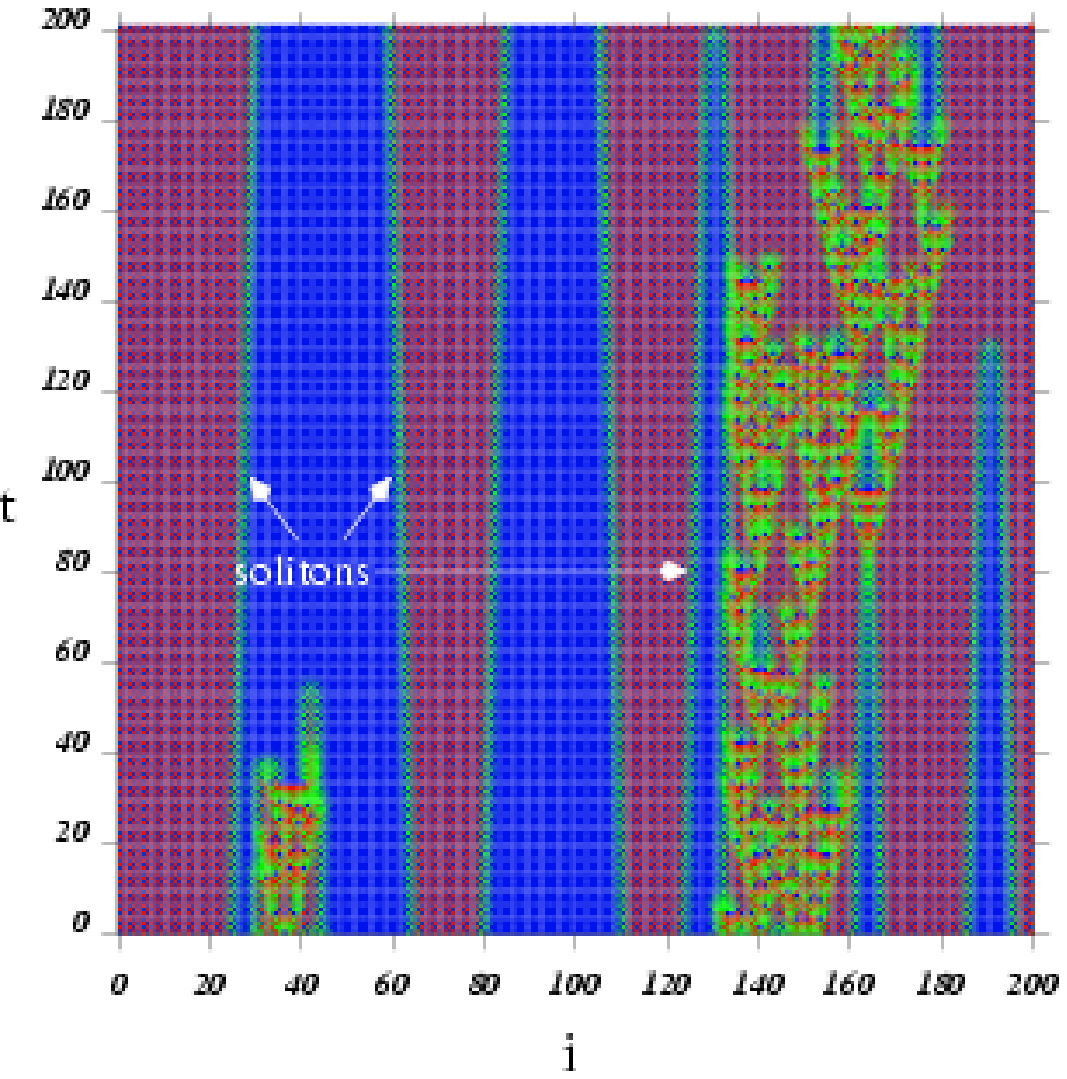}
\vspace{-.9cm}
}
\\
(c)&(d)\\
{\vspace{-.9cm}
\hspace{-.5in}\includegraphics[scale=.8]{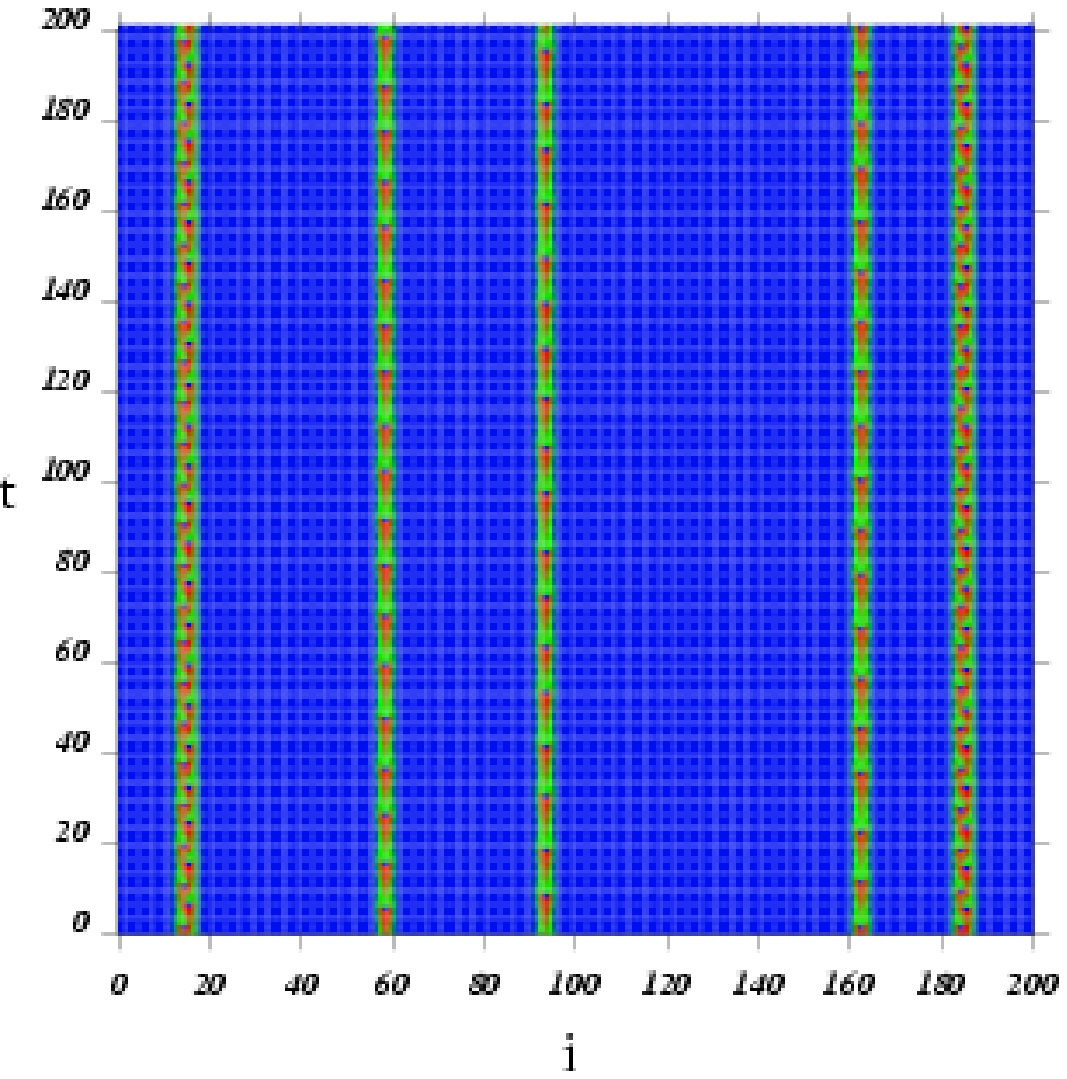}
}
&
{\hspace{-.5cm}\includegraphics[scale=.8]{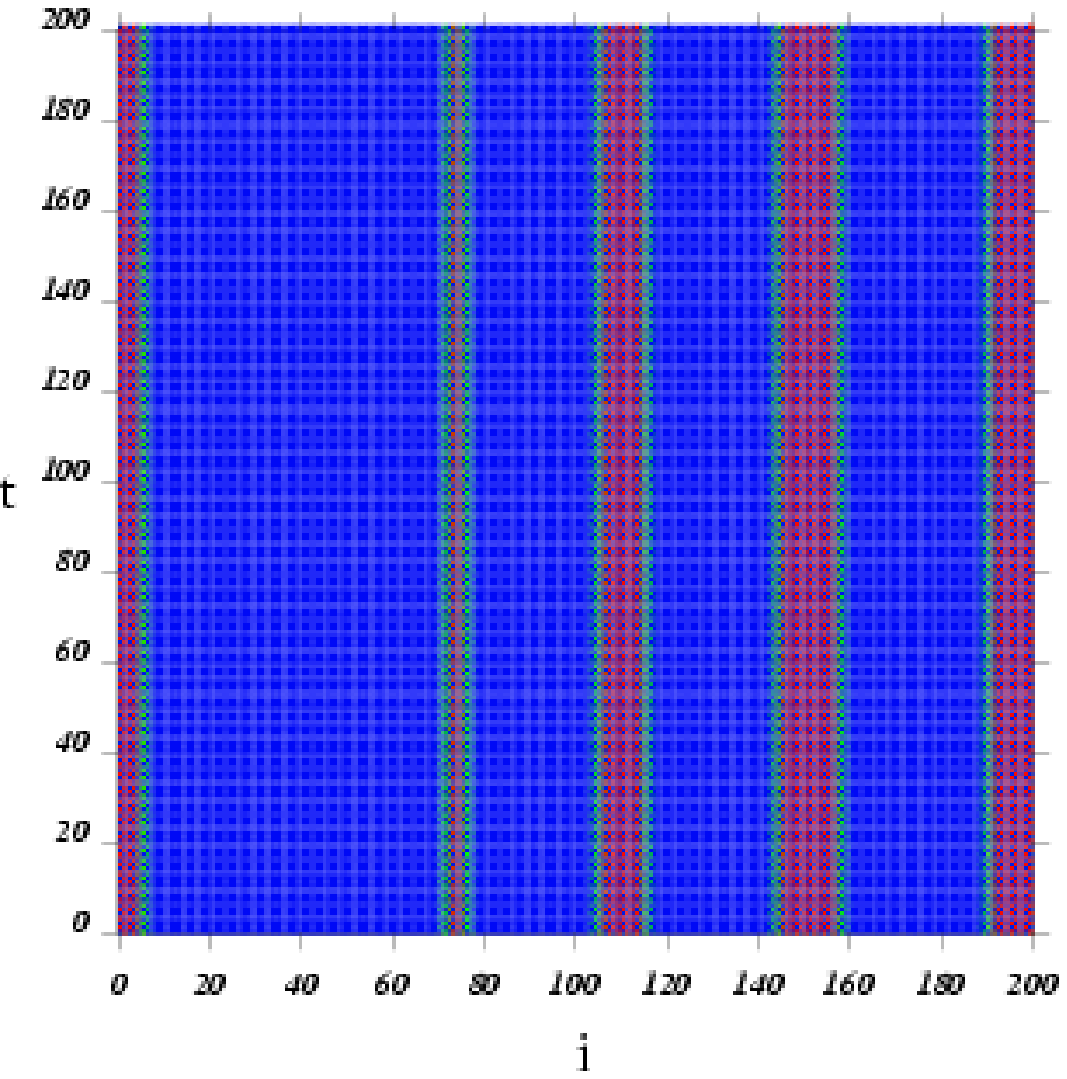}
\vspace{-1.4cm}
}\\
\end{tabular}
\end{center}
\caption{ shows the space time plots of the different types of STI seen 
in the phase diagram. The lattice index $i$ is along the x-axis and the
time index $t$ is along the y-axis. The space time plots show (a) STI with synchronized laminar state interspersed with turbulent bursts seen at $\Omega=0.06, \epsilon=0.7928$. (b) STI with TW laminar state and turbulent bursts with solitons seen at  $\Omega=0.037, \epsilon=0.937$. (c) SI with synchronized laminar state and quasi-periodic bursts seen at $\Omega=0.031, \epsilon=0.42$. (d) SI with synchronized laminar state and TW bursts observed at $\Omega=0.019, \epsilon=0.9616$.  \label{stplot}}
\end{figure}

The phase diagram obtained using random initial conditions is shown in Figure \ref{pd} \cite{zjng}. Spatially synchronized, temporally frozen solutions, where the variables $x_i(t)$ take the value $x_i(t)=x^{\star}=\frac{1}{2\pi} \sin^{-1}\frac{2\pi\Omega}{K}$ for all $i=1,\ldots,N$, and for all $t$, are marked by dots in Figure \ref{pd}. These solutions are seen over a large section of the phase diagram and are stable against perturbations. Cluster solutions, in which $x_i(t)=x_j(t)$ for all $i,j$ belonging to a particular cluster, are identified by plus signs (+) in the phase diagram. Regimes of spatiotemporal intermittency of various kinds are seen near the bifurcation boundary of the synchronized solutions. The various types of STI seen are:
\begin{enumerate}
\item[i.] STI of the type in which the laminar state is the synchronized fixed point $x^{\star}$ defined earlier, and the turbulent state takes all other values other than $x^{\star}$ in the $[0,1]$ interval, is seen at points marked with diamonds ($\Diamond$) in Figure \ref{pd}. The space-time plot is shown in Figure \ref{stplot}(a). This type of STI belongs to the directed percolation universality class.
\item[ii.] STI with TW laminar state interspersed with turbulent bursts
is seen at points marked with boxes ($\Box$) in Figure \ref{pd}. The
space-time plot of this type of solutions is shown in Figure
\ref{stplot}(b). Coherent structures traveling in space and time are
seen in these solutions. Such structures have also been seen in the
Chat\'e - Manneville CML and have been called solitons in the
literature. 
\end{enumerate}

\begin{enumerate}
\item[iii.] Spatial intermittency with synchronized laminar state and quasi-periodic  bursts are seen at parameters marked with triangles ($\triangle$) in the phase diagram. The space-time plot of this type of solution is shown in Figure \ref{stplot}(c). 
\item[iv.] Spatial intermittency with synchronized laminar state and traveling wave (TW) bursts are seen at points marked with crosses ($\times $) in the phase diagram.  The space-time plot is shown in Figure \ref{stplot}(d). SI with synchronized laminar state and period-5 bursts are seen at points marked with asterisks ($\ast$) in the phase diagram. 
\end{enumerate}

The identification of the universality classes of the different types of
intermittency seen in this system has been partially carried out
earlier. STI with synchronized laminar states and turbulent bursts has
been clearly established to belong to the directed percolation (DP) class
\cite{jan,zjngpre}. However, the other types of intermittency seen in the parameter 
space do not belong to the DP class. We analyse these in the next
section.

\section{The universality classes in the system}

It is interesting to note that the system under study exhibits
spatiotemporal intermittency belonging to distinct universality
class at
different values of the parameters. The two distinct classes obtained so
far are regimes of STI which belong to the DP class and regimes of SI
which do not belong to the DP class. We discuss each of these in further
detail.
\subsection{STI of the DP type}

\begin{table}[!b]
\begin{center}
\begin{tabular}{ll|cccccc|ccccc}
\hline
\multicolumn{11}{c}{\bf Static and dynamic scaling exponents for the STI of the DP class}\\
\hline
\multirow{2}{*}{~~ $\Omega$~~}&\multirow{2}{*}{ ~~$\epsilon_c(\Omega)$~~}&\multicolumn{6}{c}{ Bulk exponents }	&\multicolumn{3}{|c}{ Spreading Exponents}\\
\cline{3-11}
 & &  $z$ &  $\beta/{\nu z} $ &  $\beta$ &  $\nu$ &  $\eta'$ & { $\zeta$} & $\eta$	&$\delta$	&$z_s$\\
\hline
~~0.060~~	&~~0.7928~~	&~~1.59~~	&~~0.17~~	&~~0.293~~	&~~1.1~~	&~~1.51~~	&~~1.68~~&~~0.315~~ &~~0.16~~	&~~1.26~~	\\
\hline
~~0.073~~	&~~0.4664~~	&~~1.58~~	&~~0.16~~	&~~0.273~~	&~~1.1~~	&~~1.50~~	&~~1.65~~&~~0.308~~ &~~0.17~~	&~~1.27~~	\\
\hline
~~0.065~~	&~~0.34949~~	&~~1.59~~	&~~0.16~~	&~~0.273~~	&~~1.1~~	&~~1.50~~	&~~1.66~~&~~0.303~~ &~~0.16~~	&~~1.27~~	\\
\hline
~~~~Error	&bars~~	&~~0.01~~	&~~0.01~~	&~~-~~		
&~~-~~		&~~0.01~~	&~~0.01~~&~~0.001~~ &~~0.01~~	&~~0.01~~	\\
\hline
\multicolumn{2}{c|}{\bf DP}& {\bf 1.58}	& {\bf 0.16}	& {\bf 0.28}	& {\bf 1.1}	& {\bf 1.51}	&{\bf 1.67} & {\bf 0.313}	&{\bf 0.16}	&{\bf 1.26}		\\
 \hline
\end{tabular}
\caption{ The static and dynamic exponents obtained  in the DP regime at
three of the points marked with diamonds ($\Diamond$) in the phase diagram are shown
in the above table. The universal DP exponents are listed in the last
row. The exponents, $\beta$ and $\nu$ have been calculated using the hyperscaling relations. See \cite{jan} for definitions of the DP exponents\label{dpt}.}
\end{center}
\vspace{-.1in}
\end{table}

\begin{figure}
\centering
\begin{tabular}{c}

(a)\\
{\includegraphics[height=6.cm,width=7.5cm]{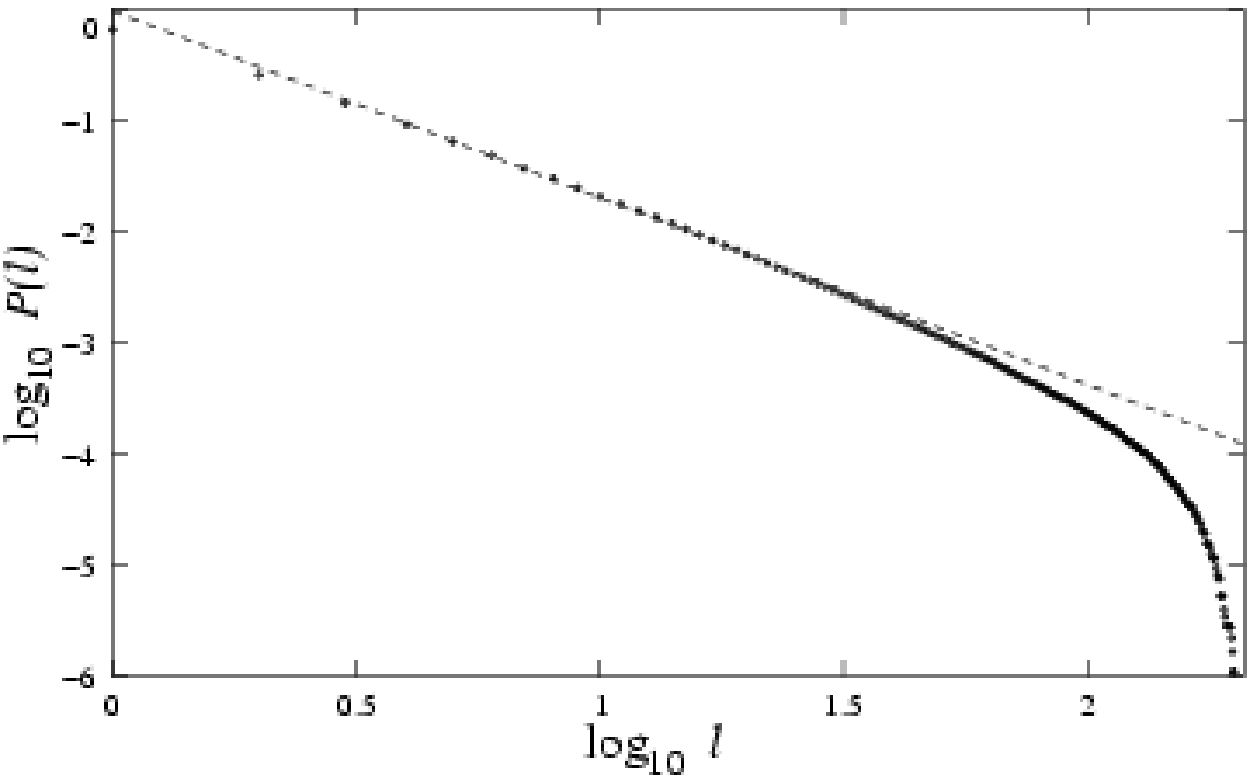}}\\

\end{tabular}

\begin{tabular}{cc}

(b)&(c)\\
{\includegraphics[height=6.cm,width=7.5cm]{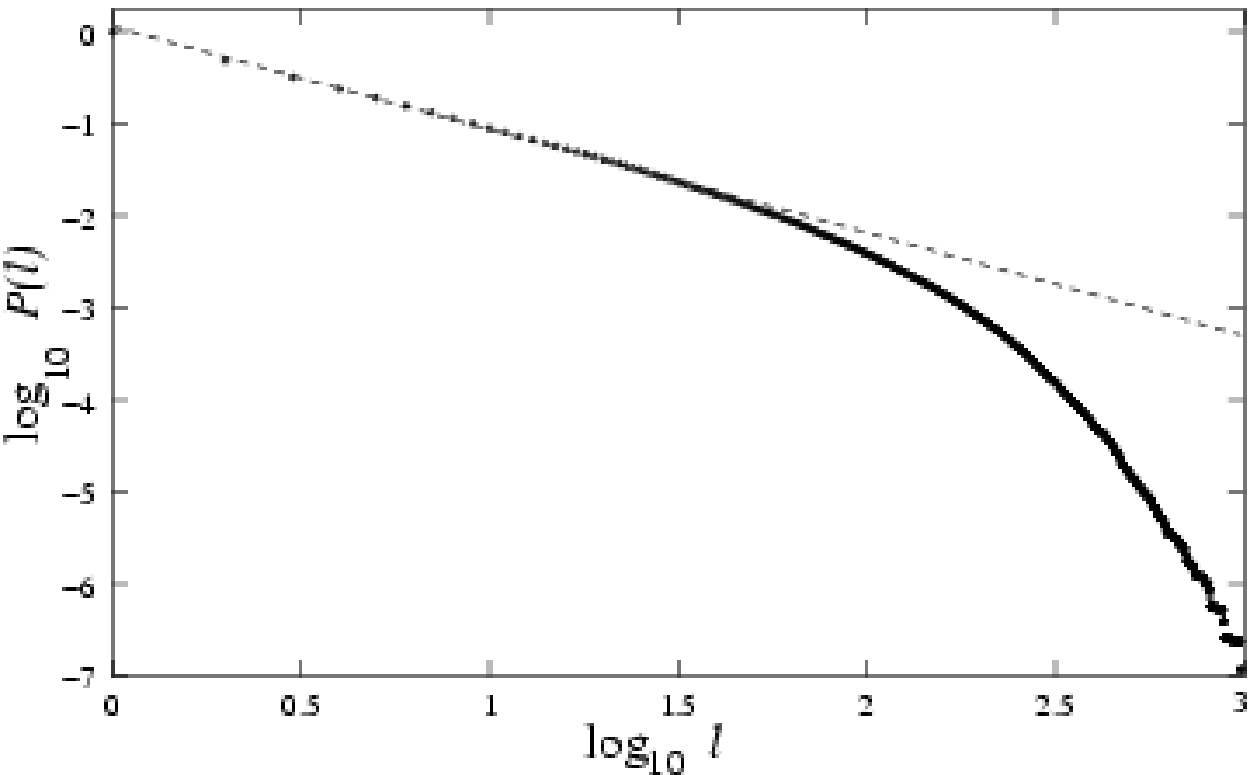}}& {\includegraphics[height=6cm,width=7.5cm]{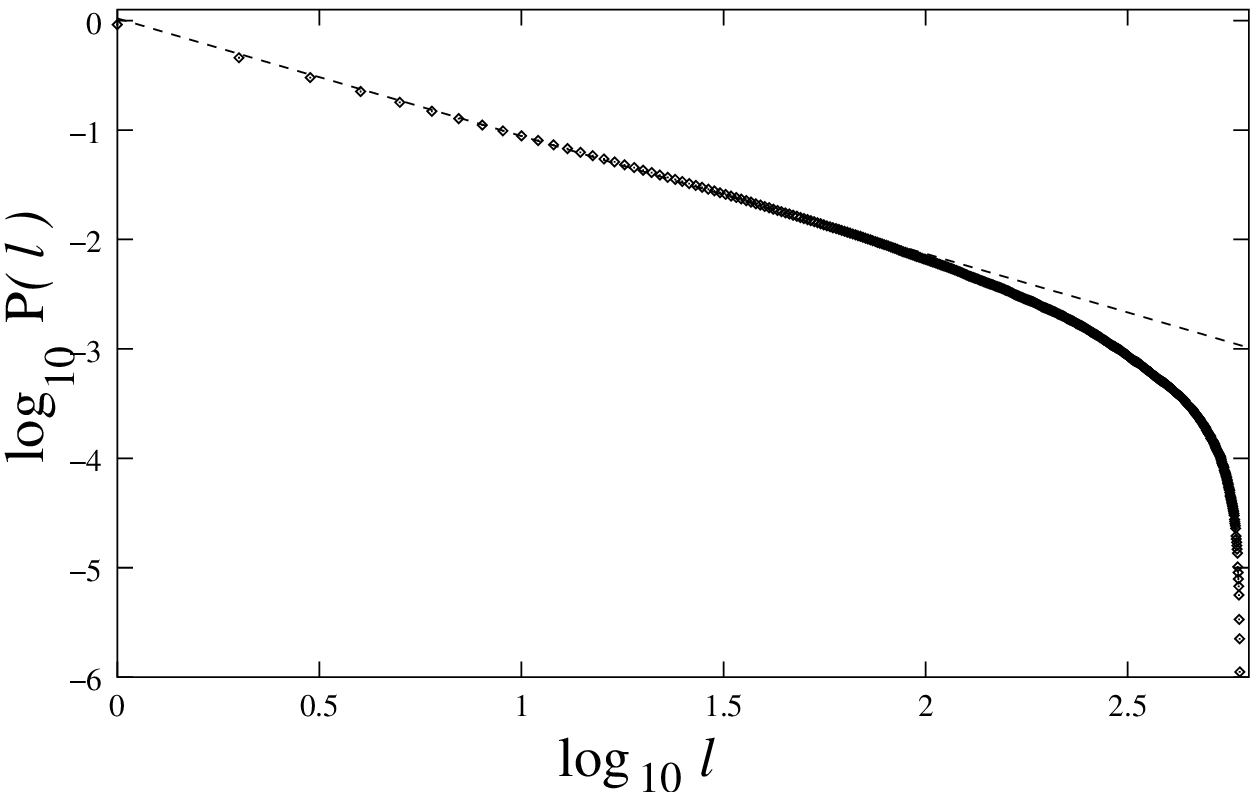}}

\end{tabular}
\caption{shows the $\log-\log$ (base 10) plot of the laminar length distribution for (a)  STI of the DP class obtained at $\Omega=0.06,\epsilon=0.7928$. The exponent $\zeta$ is 1.681. (b) SI with quasi-periodic bursts obtained at $\Omega=0.04, \epsilon=0.402$. The exponent $\zeta$ is 1.10. (c) SI with TW bursts obtained at $\Omega=0.019, \epsilon=0.9616$. The exponent obtained is 1.08. \label{lam}}
\end{figure}

It has been shown convincingly that STI with synchronized laminar state
interspersed with turbulent bursts (seen at points marked with
 $\Diamond$ in Figure \ref{pd}) belongs to the DP universality class
 \cite{jan,zjngpre}. The infective dynamics of the turbulent bursts
wherein the turbulent site either infects the adjacent laminar sites, or
dies down to the laminar state, is similar to the behaviour seen in
directed percolation \cite{stauffer}. Since no spontaneous creation of turbulent bursts
takes place, the laminar state forms the absorbing state and time acts
as the directed axis.
The entire set of static and dynamic scaling exponents obtained in this parameter regime match with the DP exponents. The exponents obtained, after averaging over $10^3$ initial conditions, at three such parameter values are listed in Table \ref{dpt}. The complete set of exponents and their definitions have been reported in \cite{jan, zjngpre}. The distribution of laminar lengths also shows a scaling behaviour of the form, $P(l)\sim l^{-\zeta}$, with an associated exponent, $\zeta\sim1.67$.  The laminar length distribution obtained, after averaging over $50$ initial conditions, at $\Omega=0.06,\epsilon=0.7928$ has been plotted in Figure \ref{lam}(a). The size of the lattice studied was $10^4$.

A clean set of DP exponents is obtained for the
STI with synchronized laminar state seen in this parameter regime as these regimes are completely free 
from the 
presence of  coherent soliton-like structures which could bring in 
long-range correlations in the system and thereby spoil the DP behaviour.
In fact, in the case of the STI with synchronized laminar state, no 
solitons have 
been observed for this model in the  range of parameters studied.
However, the STI with traveling wave laminar states seen at the
parameter values marked with boxes in the phase diagram, does show the
presence of solitons as seen    
in the space-time plot of Fig.\ref{stplot}(b). These
solitons are responsible for non-universal exponents and 
cross-over behaviour in this regime. This behaviour is discussed in detail in
section IV. In the remainder of the present section we will discuss the 
second universality class seen for the present model, viz. that of
spatial intermittency.

\subsection{Spatial Intermittency}
\begin{table}[!t]
\begin{center}
\begin{tabular}{ccccc}
\hline
\multicolumn{5}{c}{\bf Spatial intermittency with quasi-periodic bursts} \\
\hline
 {$\Omega$} & {$\epsilon$} & {$\zeta$} &{$\kappa$} & {$\omega$} \\
\hline
 ~~~~0.005~~~~ & ~~~~0.520~~~~ & ~~~~1.08 $\pm$ 0.01~~~~ & ~~~~1.07$\pm$0.01~~~~& ~~~~0.007, 0.014, 0.022~~~~\\
 ~~~~0.010~~~~ & ~~~~0.505~~~~ & ~~~~1.13 $\pm$ 0.01~~~~ & ~~~~1.06$\pm$0.01~~~~&~~~~0.027, 0.054, 0.081~~~~\\
~~~~0.015~~~~ & ~~~~0.480~~~~ & ~~~~1.11 $\pm$ 0.03~~~~ & ~~~~1.14$\pm$0.02~~~~&~~~~0.037, 0.073, 0.110~~~~\\
 ~~~~0.035~~~~ & ~~~~0.418~~~~ & ~~~~1.10 $\pm$ 0.05~~~~ & ~~~~1.43$\pm$0.03~~~~&~~~~0.110, 0.230, 0.340~~~~\\
  ~~~~0.040~~~~ & ~~~~0.402~~~~ & ~~~~1.10 $\pm$ 0.04~~~~ & ~~~~1.31$\pm$0.01~~~~&~~~~0.150, 0.230, 0.380~~~~\\
~~~~0.044~~~~ & ~~~~0.373~~~~ & ~~~~1.09 $\pm$ 0.03~~~~ & ~~~~1.18$\pm$0.03~~~~&~~~~0.060, 0.120, 0.180~~~~\\
~~~~0.059~~~~ & ~~~~0.286~~~~ & ~~~~1.16 $\pm$ 0.02~~~~ & ~~~~1.07$\pm$0.02~~~~&~~~~0.066, 0.133, 0.200~~~~\\
 \hline

\end{tabular}
\end{center}
\caption{ The table shows the laminar length distribution exponent, $\zeta$, 
calculated for SI with quasi-periodic bursts (marked by triangles ($\triangle$) in
Figure \ref{pd}). The exponent $\kappa$ is the exponent associated with the number
of gaps, $N_g(l)\sim l^{-\kappa}$ in the eigenvalue distribution, where $l$ is the
bin size chosen. The frequencies, $\omega$ inherent in the time series of the burst state are also listed and are of the form $\omega_1$, $\omega_2$, and $\omega_1+\omega_2$.
\label{si1}}
\end{table}

We now discuss 
the other type of intermittency, seen in this system viz. spatial
intermittency.
Spatial Intermittency is a distinct class of STI in which the temporal
behaviour of both the laminar and burst states is regular. The infective 
dynamics characteristic of the DP class is absent here, and the burst
states do not infect the laminar states even when they are nearest
neighbours. Spatial intermittency is a long lived phenomenon and 
the spatially intermittent state persists for time scales which are much
longer than the time scales on which the STI states die down to a
uniform laminar background. 
Two different types of spatial intermittency have been seen in this
system. In both types of SI, the laminar state is the synchronized fixed
point, $x^{\star}$ defined earlier. However, the burst states are
different and may be either quasi-periodic (marked by triangles in the
phase diagram) or periodic 
in their temporal behaviour.

\begin{enumerate}
\item {\bf SI with quasi-periodic bursts}

Spatial intermittency, in which the temporal behaviour of the burst
states is quasi-periodic in nature, has been seen at points marked with
triangles ($\triangle$) in Figure \ref{pd}. The space-time plot has been
shown in Figure \ref{stplot}(c). These burst states are non-infective in
nature, i.e. the probability of the burst state infecting the nearby
laminar state is zero. Therefore, the laminar states remain laminar
forever. Hence, after an initial transient, the order parameter of the
system, which is defined as the fraction of non-laminar sites in the
lattice, is a constant. The time series of the burst states at  
different parameter values, at a typical burst site, was studied using power spectrum analysis. The power spectrum obtained at $\Omega=0.058,\epsilon=0.291$ and $\Omega=0.0495,\epsilon=0.3178$ has been shown in Figure \ref{psts}. As can be seen from Figure \ref{psts}(a), the peaks are seen at $\omega_1$, $\omega_2$, $\omega_1 + \omega_2$ and at $m  \omega_1 + n\omega_2$. This kind of behaviour is typical of a quasi-periodic state. Hence, we confirm that a quasi-periodic burst state is seen at $\Omega=0.058,\epsilon=0.291$.

\begin{figure}
\begin{center}
\begin{tabular}{cc}
\includegraphics[height=5.5cm,width=7.5cm]{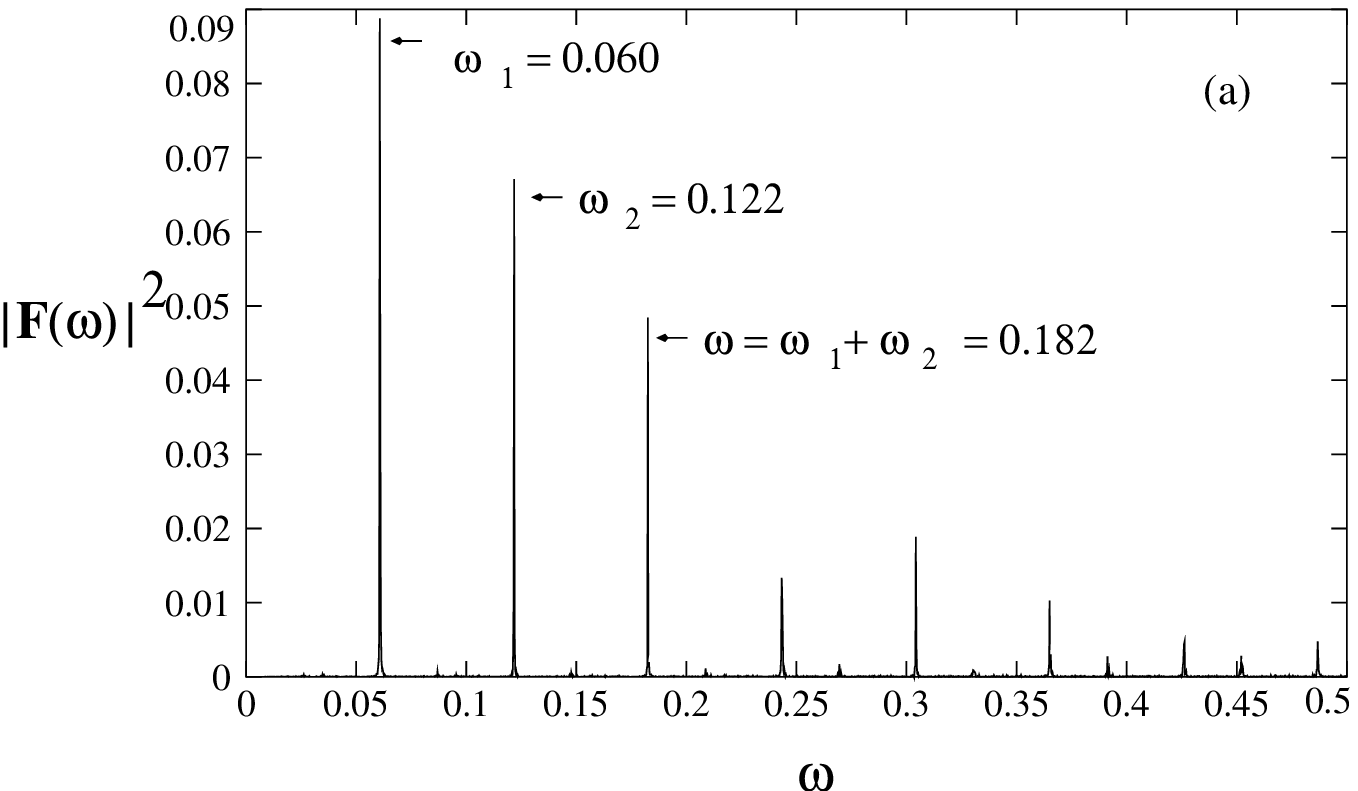} & \includegraphics[height=5.5cm,width=7.5cm]{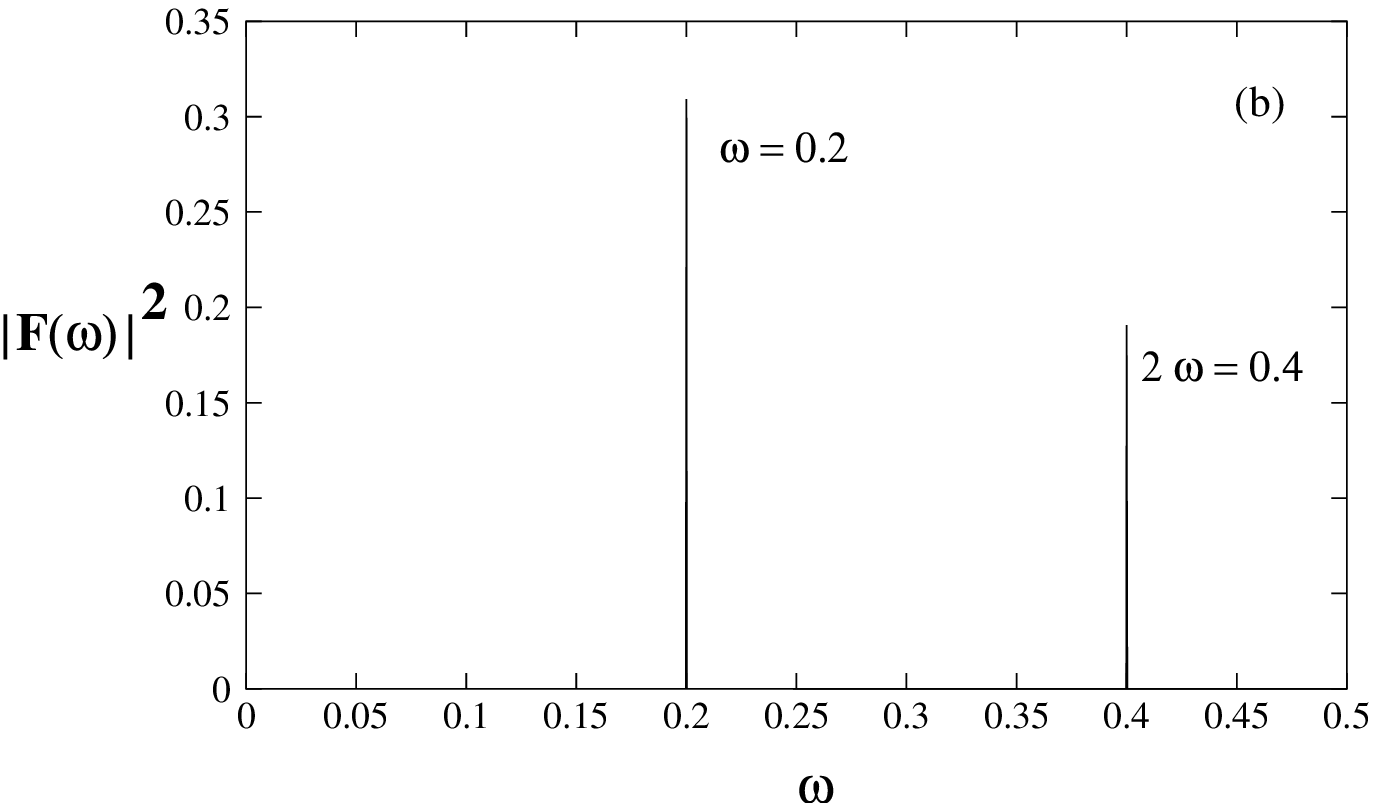}
\end{tabular}
\end{center}
\caption{shows the power spectrum , $|F(\omega)|^2$ of the time series of the burst state seen at (a) $\Omega=0.058,\epsilon=0.291$ and (b) $\Omega=0.0495,\epsilon=0.3178$. The time series shown in (a) exhibits quasi-periodic behaviour and the time series shown in (b) is periodic in nature. \label{psts}}
\end{figure}

\begin{figure}[!b]
\begin{center}
\begin{tabular}{cc}
{\hspace{-1.0cm}\includegraphics[height=5.2cm,width=6.2cm]{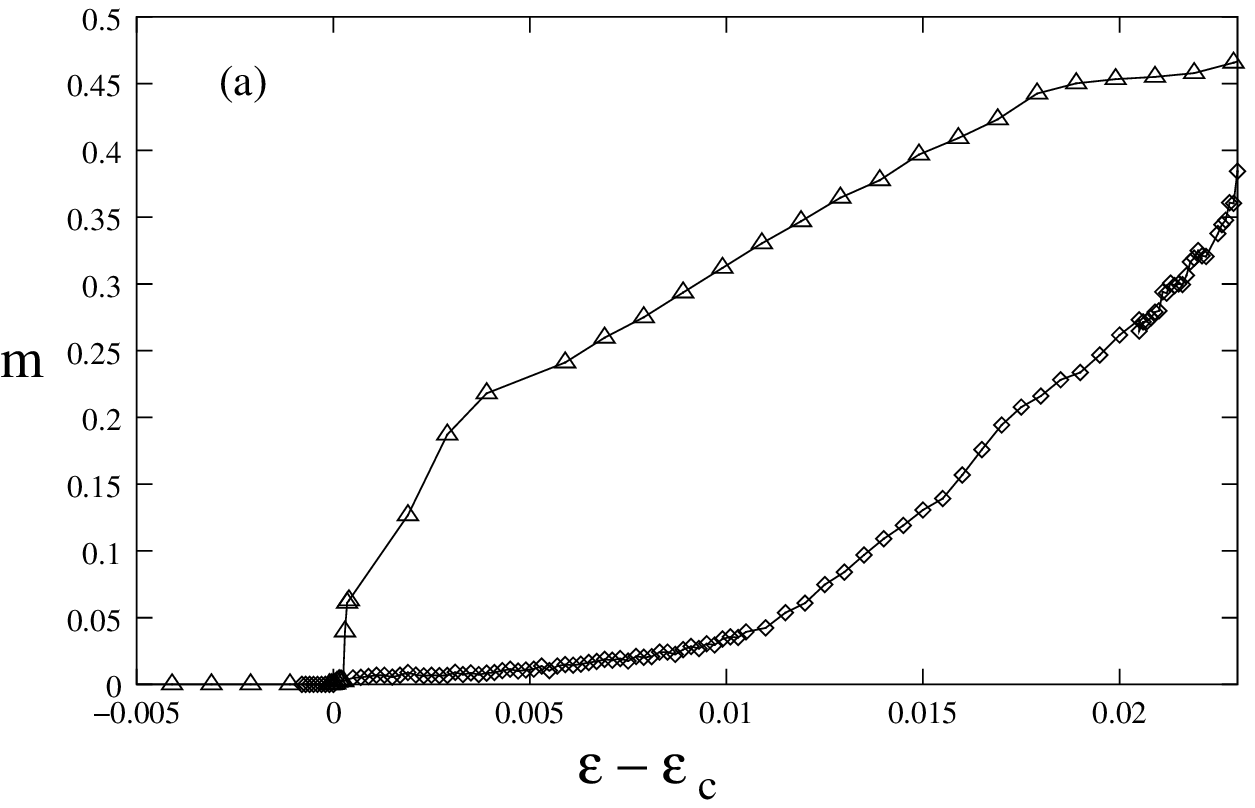}}&
{\includegraphics[height=5.2cm,width=6.2cm]{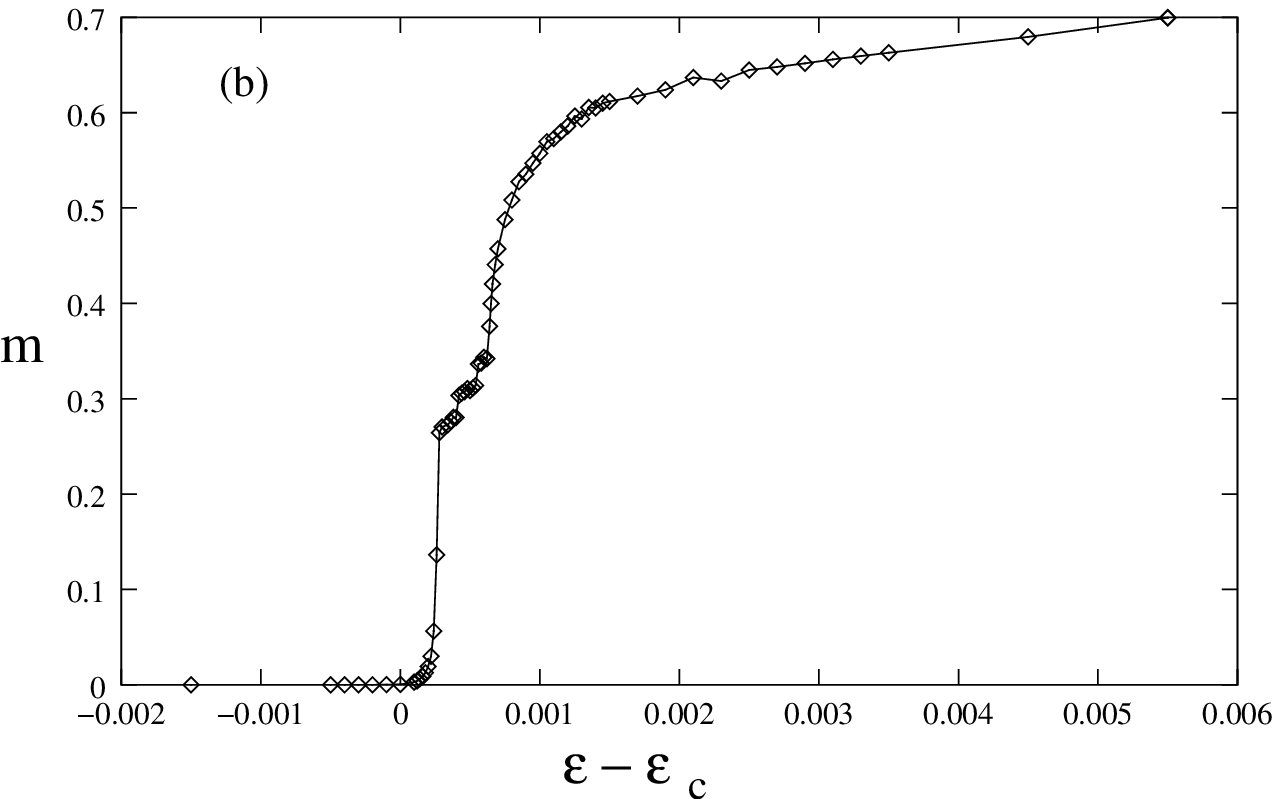}}\\
\end{tabular}
\end{center}
\vspace{-.3in}
\caption{ shows the order parameter, m vs $(\epsilon-\epsilon_c)$
plotted for SI with synchronized laminar state and quasiperiodic bursts
at $\Omega=0.031$ ($\triangle$) and STI of the DP class at $\Omega=0.06$
($\Diamond)$ in Fig. (a). Fig (b) shows m vs $(\epsilon-\epsilon_c)$
plotted for SI with TW bursts at $\Omega=0.042$. SI shows a first order
transition whereas DP class shows a second order transition. The data
collected is for a $1000$ site lattice and is averaged over $1000$
initial conditions.\label{order}}
\end{figure}

The laminar length distribution of this type of SI shows a scaling behaviour 
of the form, $P(l)\sim l^{-\zeta}$ with an associated exponent, $\zeta \sim 1.1$. The laminar length distribution at parameters $(\Omega=0.04,\epsilon=0.402)$ has been plotted in Figure \ref{lam}(b). The values of $\zeta$ obtained for this type of SI at different values of $(\Omega,\epsilon)$ have been listed in Table \ref{si1}. The scaling exponent, $\zeta$ obtained for this type of SI is clearly different from that of the DP class ($\zeta_{DP} =1.67$).

\indent~~ Secondly, the transition to SI from a completely synchronized
state is a first order transition unlike the transition to STI belonging to the directed percolation class which shows a second order transition. This can be seen in Figure \ref{order}(a) in which the order parameter of the system $m$, which is defined as the fraction of burst states in the lattice, has been plotted as a function of the coupling strength, $\epsilon$. The order parameter, $m$ increases continuously with $\epsilon$ in the case of STI of the DP class, signalling a second order transition, whereas $m$ shows a sharp jump with $\epsilon$ in the case of SI with quasi-periodic bursts, indicating that a first order transition takes place in the case of SI.
\item  {\bf SI with periodic bursts}

Spatial intermittency with periodic bursts forms the second class of SI. 
Two distinct burst periods have been observed in the phase diagram. Bursts 
of period $5$ have been seen at the points marked by asterisks ($\ast$) in the phase diagram. 

\begin{table}[!t]
\begin{center}
\begin{tabular}{ccccc}
\hline
\multicolumn{5}{c}{\bf Spatial intermittency with periodic bursts} \\
\hline
 {$\Omega$} & {$\epsilon$} & {$\zeta$} & {$\kappa$}& $\omega$\\
\hline
~~~~0.019~~~~ & ~~~~0.9616~~~~ & ~~~~1.08 $\pm$ 0.04~~~~ & ~~~~1.18$\pm$0.04~~~~&~~~~0.5~~~~\\
~~~~0.025~~~~ &  ~~~~0.9496~~~~ & ~~~~1.08 $\pm$ 0.02~~~~ &~~~~1.12$\pm$0.02~~~~& ~~~~0.5~~~~\\
~~~~0.037~~~~ & ~~~~0.9254~~~~ & ~~~~1.17 $\pm$ 0.02~~~~ &~~~~1.07$\pm$0.01~~~~&~~~~0.5~~~~\\
~~~~0.042~~~~ & ~~~~0.9148~~~~ & ~~~~1.13 $\pm$ 0.02~~~~ &~~~~1.10$\pm$0.04~~~~& ~~~~0.5~~~~\\
~~~~0.047~~~~ & ~~~~0.3360~~~~ & ~~~~1.13 $\pm$ 0.02~~~~ &~~~~1.02$\pm$0.01~~~~& ~~~~0.2, 0.4~~~~\\
~~~~0.0495~~~~ & ~~~~0.3178~~~~ & ~~~~1.15 $\pm$ 0.04~~~~ &~~~~1.03$\pm$0.02~~~~& ~~~~0.2, 0.4~~~~\\
~~~~0.054~~~~ & ~~~~0.2936~~~~ & ~~~~1.17 $\pm$ 0.03~~~~ &~~~~1.02$\pm$0.02~~~~& ~~~~0.2, 0.4~~~~\\
\hline
\end{tabular}
\end{center}
\caption{ The table shows the laminar length distribution exponent, $\zeta$ calculated for SI with periodic bursts (marked by crosses ($\times$) and asterisks ($\ast$) in Figure \ref{pd}). The exponent $\kappa$ is the exponent associated with the number of gaps, $N_g(l)\sim l^{-\kappa}$ in the eigenvalue distribution, where $l$ is the bin size chosen. The frequencies, $\omega$ inherent in the time series of the burst state are also listed. \label{si2}}
\end{table}

 Figure \ref{psts}(b) shows that peaks are seen in the power spectrum of the burst state time series at $\omega=0.2$ and higher harmonics. This confirms that  the burst states have  period $5$ at $\Omega=0.0495,\epsilon=0.3178$, which is one of the points marked by asterisks in the phase diagram.

Bursts of spatial period two, temporal period two, of the traveling wave (TW) type are seen at the points marked by crosses ($\times $) in the phase diagram.  
The laminar state in both cases is  the spatiotemporally synchronized
fixed point, $x^{\star}$. The space-time plot of these SI with TW burst
solutions at $\Omega=0.019,\epsilon=0.9616$ is shown in Figure
\ref{stplot}(d). The bursts are non-infective in nature in this type of
SI as well. The scaling exponent, $\zeta$ associated with the laminar length distribution at different parameter values in this regime have been listed in Table \ref{si2} for both TW and period $5$ bursts. The laminar length distribution exponent obtained in this regime is $\zeta\sim 1.1$. The transition to SI with TW burst state from a spatiotemporally synchronized state is also a first order transition as has been shown by the abrupt jump in the order parameter, $m$ with change in the coupling strength, $\epsilon$ (Figure \ref{order}(b)).

It is thus clear that SI does not belong to the DP universality class. The scaling exponent $\zeta=1.1$ for laminar lengths for the SI is distinctly different from the DP exponent $\zeta=1.67$. We note, however, that the nature of the bursts, viz. periodic or quasi-periodic, has no effect on the value of the exponent $\zeta$. We hence conclude that SI with periodic as well as quasi-periodic bursts belong to the same class. 
 A similar value of the laminar length distribution exponent ($\zeta\sim 1.1$) has been reported for spatial intermittency in the inhomogenuously coupled logistic map lattice \cite{ash}. 
Thus spatial intermittency appears to constitute a distinct universality class of the non-DP type.

\end{enumerate}

Therefore, two distinct universality classes of spatiotemporal
intermittency, viz. directed percolation and spatial intermittency, are
seen in the coupled sine circle map lattice in different regions of the
parameter space. The reasons for the appearance of these two distinct
classes  may lie in the long-range correlations in the system at 
different parameter values.

\section{The role of solitons in STI with TW laminar state}

As mentioned in Section III.A, in addition to spatiotemporal intermittency with synchronized laminar states, the phase diagram of our model also shows spatiotemporal intermittency with TW laminar states and turbulent bursts at the points marked by boxes in the phase diagram. The lattice dies down to the absorbing TW laminar state from random initial conditions asymptotically. The STI with TW laminar states seen in this model appears as a result of a tangent-period doubling bifurcation from the SI with TW bursts as can be seen from Table \ref{evaltw}. 

\begin{table}[!t]
\begin{tabular}{cccc}
\hline
\multicolumn{4}{c}{\bf Bifurcations from SI with TW bursts}\\
\hline
$\Omega$ & $\epsilon$ & Eigenvalues & Type of bifurcation\\
\hline
~~~~0.0100~~~~ & ~~~~0.982~~~~ & 1.685,~ -1.618 & TP\\
~~~~0.0210~~~~ & ~~~~0.960~~~~ & 1.361,~ -1.222 & TP\\
~~~~0.0305~~~~ & ~~~~0.943~~~~ & 1.752,~ -1.535 & TP\\
~~~~0.0410~~~~ & ~~~~0.920~~~~ & 1.623,~ -1.309 & TP\\
\hline
\end{tabular}
\caption{shows the largest positive and largest negative eigenvalues observed when SI with TW bursts (marked with crosses in the phase diagram) bifurcates to STI with TW laminar states (marked with boxes). The solution changes through a tangent-period doubling (TP) bifurcation.\label{evaltw}}
\end{table}

\begin{table}[!b]
\begin{tabular}{cc|cc}
\hline
\multicolumn{4}{c}{\bf Laminar length exponents in the solitonic regime}\\
\hline
\multicolumn{2}{c|}{$\Omega=0.035$}&\multicolumn{2}{c}{$\Omega=0.037$}\\
\hline
~~~$\epsilon$~~~ & $\zeta$ & $~~~\epsilon~~~~$ & $\zeta$\\
\hline
~~~~0.933~~~~ & ~~~~1.53 $\pm$ 0.01~~~~ & ~~~~0.930~~~~ & ~~~~1.50 $\pm$ 0.02~~~~\\
~~~~0.943~~~~ & ~~~~1.40 $\pm$ 0.01~~~~ & ~~~~0.937~~~~ & ~~~~1.31 $\pm$ 0.05~~~~\\
~~~~0.950~~~~ & ~~~~1.17 $\pm$ 0.01~~~~ & ~~~~0.950~~~~ & ~~~~1.17 $\pm$ 0.01~~~~\\
~~~~0.962~~~~ & ~~~~1.02 $\pm$ 0.01~~~~ & ~~~~0.962~~~~ & ~~~~1.02 $\pm$ 0.00~~~~\\
\hline
\end{tabular}
\caption{ shows the laminar length distribution exponent, $\zeta$ obtained for different values of the coupling strength, $\epsilon$ and for different $\Omega$'s in the STI with TW laminar state and turbulent bursts regime. \label{coexpts}}
\end{table}

\begin{figure}[!t]

\begin{center}
\includegraphics[height=7cm,width=9.0cm]{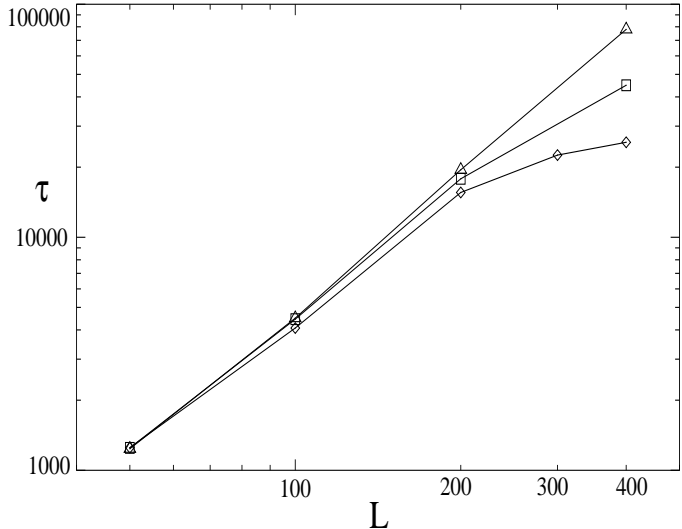}
\end{center}
\vspace{-.4in}
\caption{ shows the log-log (base 10) plot of escape time, $\tau$
plotted as a function of the lattice size, $L$ at $\Omega=0.037$ and at
$\epsilon=0.937 ~(\Diamond), 0.938 ~(\Box)$, and at $\epsilon=0.939
~(\triangle)$. The data is averaged over $1000$ initial conditions. \label{etime}}
\end{figure}

Immediately after the bifurcation, apart from turbulent states, coherent
structures, which have been called solitons, are seen in the TW laminar background. These structures have been marked in the space-time plot of this type of STI in Figure \ref{stplot}(b). 
The solitons travel through the lattice with a velocity, $v=1/{t'}$ such that for a right moving soliton, $x_{i}^t=x_{i+1}^{t+t'}$, and $x_i^t=x_{i-1}^{t+t'}$ for a left moving soliton. Here, $i$ and $t$ are the site and time indices respectively. In this model, the left and right moving solitons occur in pairs and hence, they annihilate each other.  When these solitons collide, they either die down to the TW laminar state or give rise to turbulent bursts. Such coherent structures have been seen earlier in the Chat\'e Manneville CML \cite{grassberger}, where these solitons were responsible for spoiling the DP behaviour and were even capable of changing the order of the transition. 
In the case of the sine circle map CML as well, the solitons seen in the STI with TW laminar state are responsible for non-universal exponents with values varying from $1.0$ to $1.5$ in the distribution of laminar lengths (Table \ref{coexpts}). Additionally, the   escape time, $\tau$ which is defined as the time taken for the lattice to relax to a completely laminar state, starting from random initial conditions, does not show power-law scaling as a function of $L$ for any value of the parameters (See Fig. \ref{etime}).

\begin{figure}[!t]
\begin{center}
\begin{tabular}{cc}
\hspace{-.5in}\includegraphics[height=7.0cm,width=9.0cm]{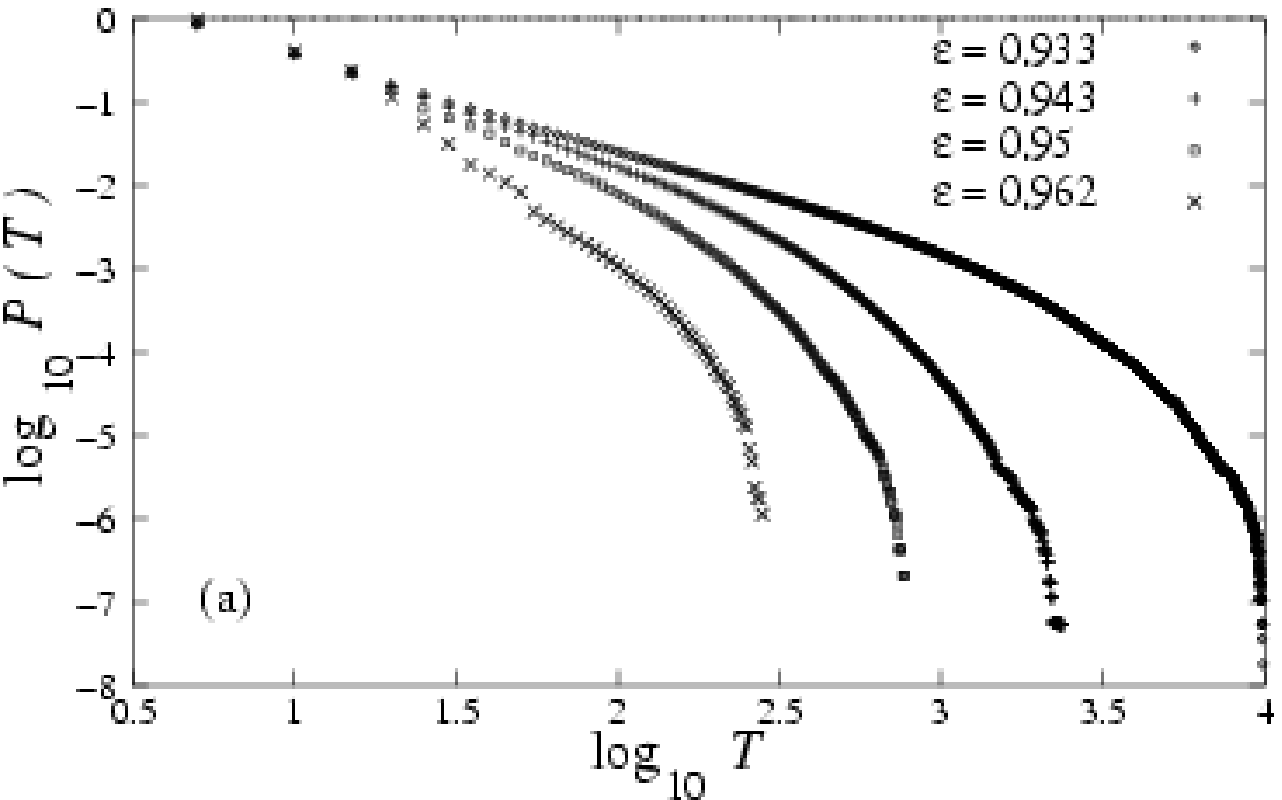}&
\includegraphics[height=7.0cm,width=9.0cm]{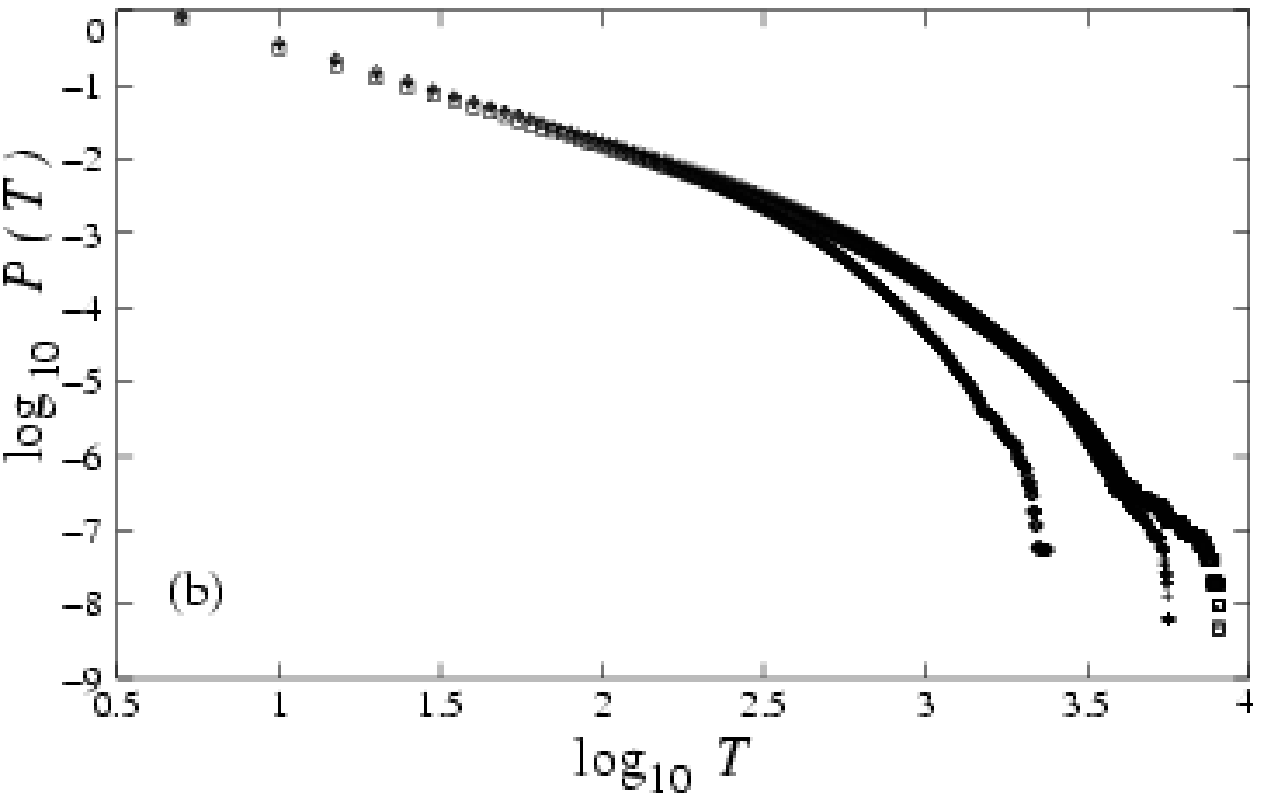}
\end{tabular}
\vspace{-.2in}
\includegraphics[height=7.0cm,width=9.0cm]{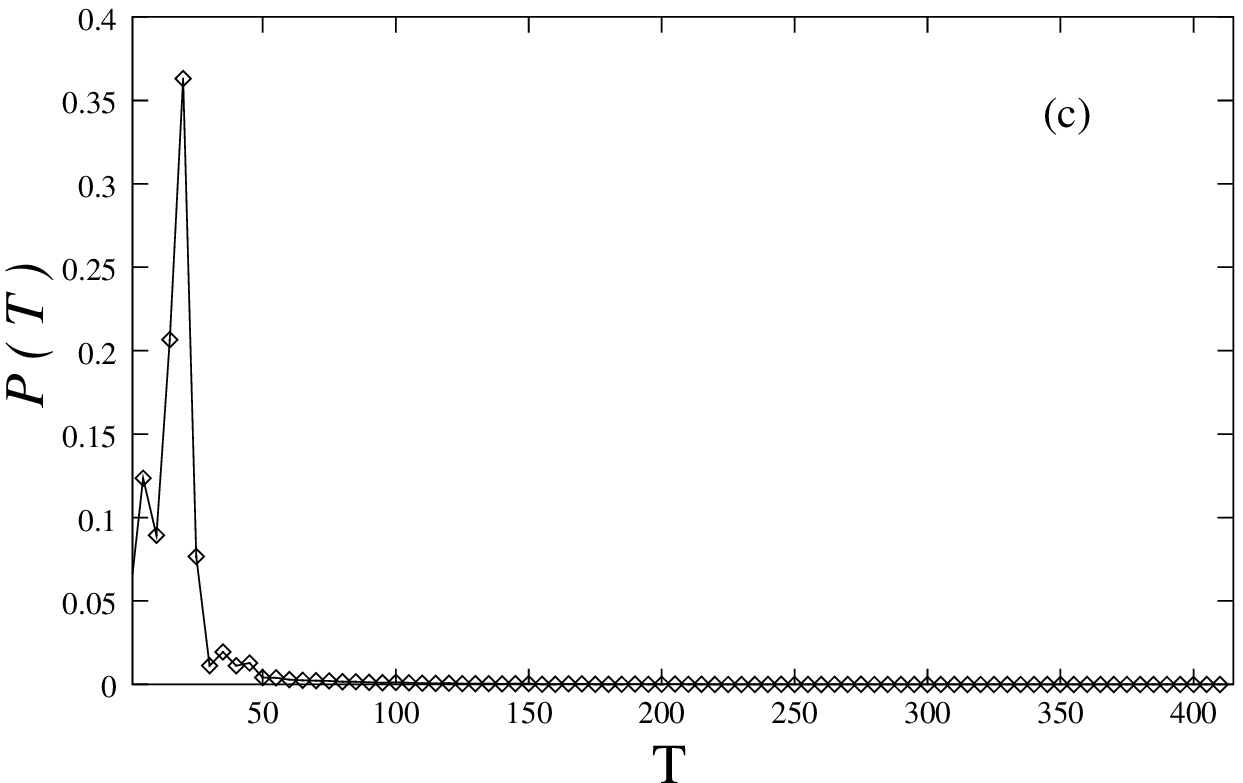}
\vspace{-.2in}
\end{center}
\caption{ (a) shows the log-log (base10) plot of distribution of soliton
lifetimes at $\Omega=0.035$ and at $\epsilon=0.933 ~(\zeta=1.53)$,
$\epsilon=0.943~ (\zeta=1.40)$, $\epsilon=0.95 ~(\zeta=1.14)$ and at
$\epsilon=0.962 ~(\zeta=1.02)$. The exponent, $\zeta$ depends on the
soliton lifetimes. (b) shows the log-log (base10) plot of the soliton
lifetime distribution at $\Omega=0.035$, $\epsilon=0.943$ (diamonds);
$\Omega=0.037, \epsilon=0.937$ (pluses) and at $\Omega=0.04,
\epsilon=0.9323$ (boxes). The laminar length distribution exponent,
$\zeta$ is $\sim 1.3$ at all these points. Hence, the soliton lifetimes
collapse on to each other. (c) shows the soliton lifetime distribution
seen in the short soliton lifetime regime at $\Omega=0.037,
\epsilon=0.962$. A peak is seen at the characteristic timescale,
$\tau_c=20$. The distributions have been obtained after evolving  a
lattice of size $500$ over $20, 000$ time steps and $50$ initial conditions.\label{soldistn}}
\end{figure}

The soliton life-times and velocities depend on the coupling strength $\epsilon$ and $\Omega$.
As the coupling strength, $\epsilon$ increases, the velocity, $v$ of the solitons is found to increase. Therefore, solitons with larger velocities collide earlier with each other, and hence have shorter lifetimes.  
The distribution of soliton life-times shows a peak in the short
life-time regime indicating the presence of a characteristic soliton life-time $\tau_c$.
However, the tail of the distribution falls off with power-law behaviour
with an exponent $2.84$.  For low values of $\epsilon$, where the
soliton velocities are smaller, there is no peak or characteristic
life-time in the distribution, and the entire distribution of soliton
life-times scales as a power-law with exponent $\sim 1.1$ (See Fig. \ref{soldistn}).
 It is seen that the exponent $\zeta$ for the laminar lengths decreases as the 
life-times decrease and the turbulent spreading in the lattice decreases. 

\begin{figure}[!t]
\begin{center}
\begin{tabular}{cc}
\hspace{-.5in}\includegraphics[height=7.0cm,width=9.0cm]{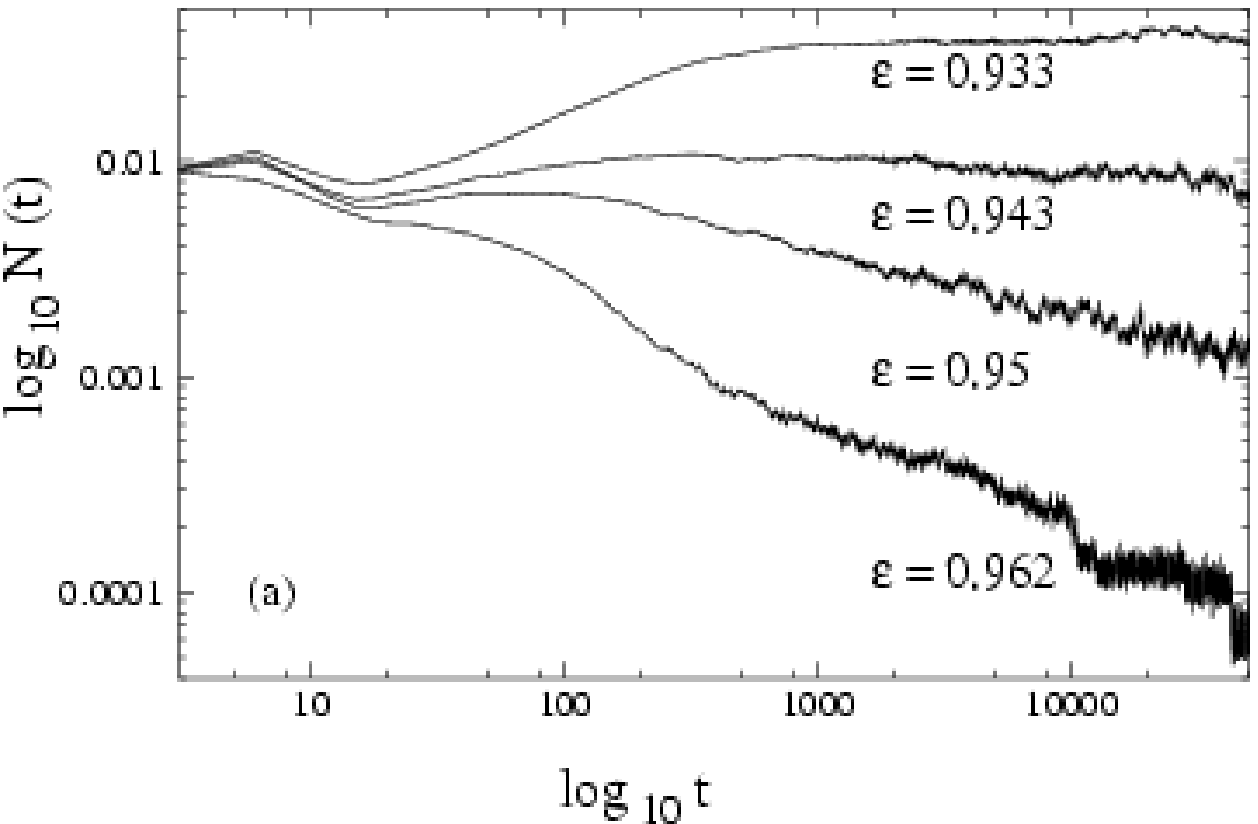}&
\includegraphics[height=7.0cm,width=9.0cm]{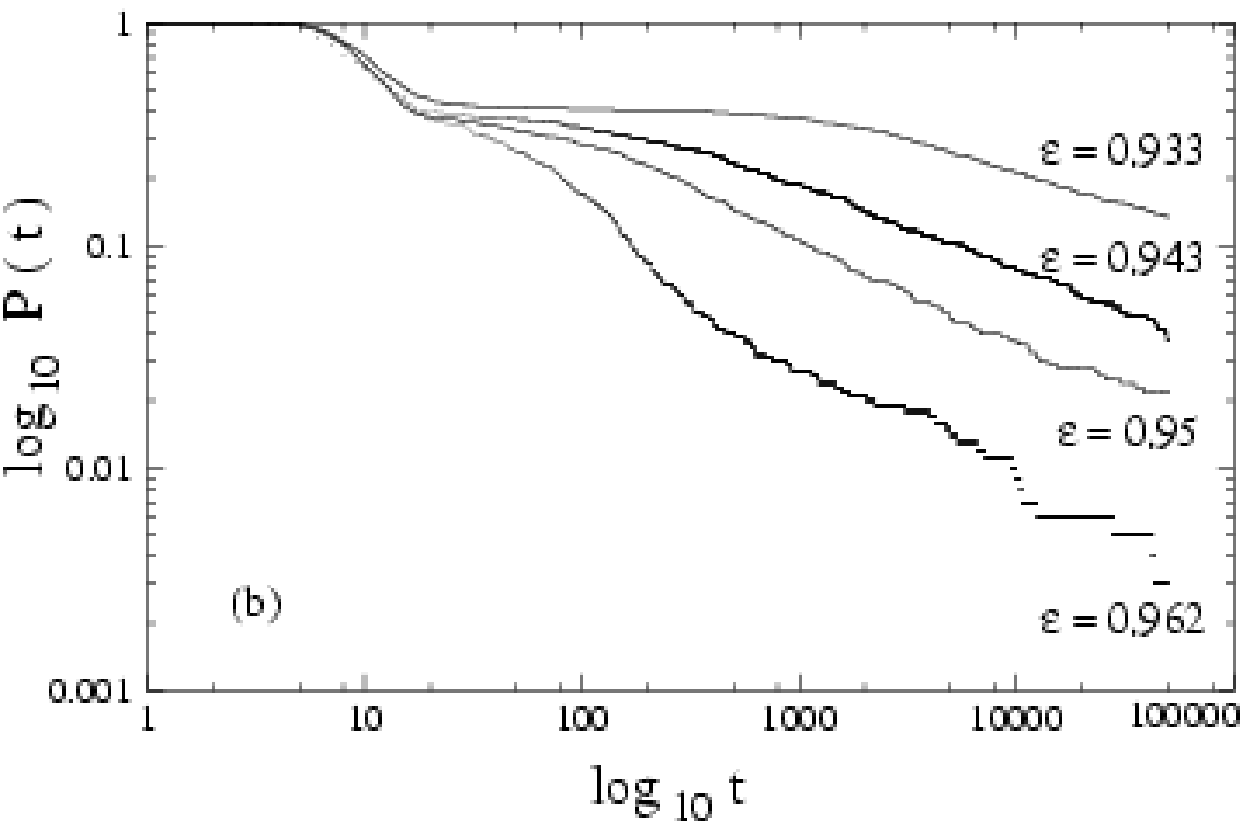}
\end{tabular}
\vspace{-.5in}
\end{center}
\caption{shows the log-log plot of (a) the fraction of turbulent sites, $N(t)$ plotted as a function of $t$ at $\Omega=0.035$, and (b) the survival probability, $P(t)$ plotted as a function of $t$ at $\Omega=0.035$. The extent of spreading decreases with increasing coupling strength, $\epsilon$ or equivalently with decreasing soliton lifetimes. \label{sprexpt}}

\end{figure}

\begin{table}[!b]
\begin{tabular}{ccccccc}
\hline
\multicolumn{7}{c}{\bf Soliton lifetimes in STI with TW laminar state}\\
\hline
Regime &$\Omega$ & $\epsilon$ & $\zeta$ & $\tau_{c}$ & $T_{max}$ &$\mu$\\
\hline
\multirow{3}{*}{~~Long soliton~~} &\multirow{2}{*}{~~~~0.035~~~~}&~~~~0.933~~~~&~~~~1.53~~~~&~~~~-~~~~&~~~~19010~~~~&~~~~1.14~~~~\\
&   &~~~~0.943~~~~&~~~~1.40~~~~&~~~~-~~~~&~~~~2481~~~~&~~~~1.35~~~~\\
\cline{2-7}
 ~~lifetimes~~&\multirow{2}{*}{~~~~0.037~~~~}&~~~~0.930~~~~&~~~~1.50~~~~&~~~~-~~~~&~~~~16961~~~~&~~~~1.18~~~~\\
 &  &~~~~0.937~~~~&~~~~1.31~~~~&~~~~-~~~~&~~~~5782~~~~&~~~~1.31~~~~\\
\hline
{~~Short soliton~~} & ~~~~0.035~~~~&~~~~0.962~~~~&~~~~1.02~~~~&~~~~15~~~~&~~~~305~~~~&~~~~2.84~~~~\\
~~lifetimes~~&~~~~0.037~~~~&~~~~0.962~~~~&~~~~1.02~~~~&~~~~20~~~~&~~~~413~~~~&~~~~2.88~~~~\\
\hline
\end{tabular}
\caption{ shows the exponents obtained in the short and long soliton lifetime regimes. Here, $\zeta$ is the laminar length distribution exponent. A characteristic timescale, $\tau_c$, is seen in the short soliton lifetime regime. $T_{max}$ is the largest soliton lifetime observed and $\mu$ is the exponent associated with the soliton lifetime distribution.\label{soltime}}
\end{table}
The spreading dynamics in this type of STI was studied by introducing a
cluster of turbulent seeds in a completely absorbing background. The two
dynamic quantities (i) $N(t)$  the fraction of turbulent sites in the lattice at
a time $t$,  and (ii) the survival probability, $P(t)$, which is defined
as the fraction of initial conditions at time $t$ which show a non-zero number of active sites, were studied at $\Omega=0.035$ and $\epsilon=0.933, 0.943, 0.95, $ and $0.962$. These have been plotted in Figure \ref{sprexpt}(a) and (b) respectively. It can be seen from the figure that the fraction of turbulent sites, $N(t)$ at a given time, $t$ decreases as the coupling strength, $\epsilon$ is increased. We see a similar decrease in the fraction of initial conditions which survive, $P(t)$ with increase in $\epsilon$. 
The data is averaged over $1000$ initial conditions.

Therefore, we can conclude that the extent of spreading in the lattice
decreases as the lifetime of the soliton decreases (with increase in
$\epsilon$). Since the distribution of laminar lengths is an indirect
measure of the spreading in the lattice, we see that the varying average soliton
lifetimes influence the distribution of laminar lengths. Therefore, the
solitons seen in this regime are responsible for non-universal exponents
here. Conversely, the soliton lifetime distributions have been plotted in Figure \ref{soldistn} (b) for parameters where  the exponents $\zeta$ for the laminar length
distributions take similar values. The soliton lifetime distributions collapse over each other as expected. 

We note again that the STI with TW laminar state shows no soliton free regime, 
and the DP regime where the laminar state is the synchronized state is completely soliton free. Hence, no direct comparison of the exponents of the STI with synchronized laminar state and STI with TW laminar state is possible at present. 

\section{Dynamic characterisers}
\begin{figure}[!t]
\begin{center}
\begin{tabular}{cc}
\includegraphics[height=6.5cm,width=8cm]{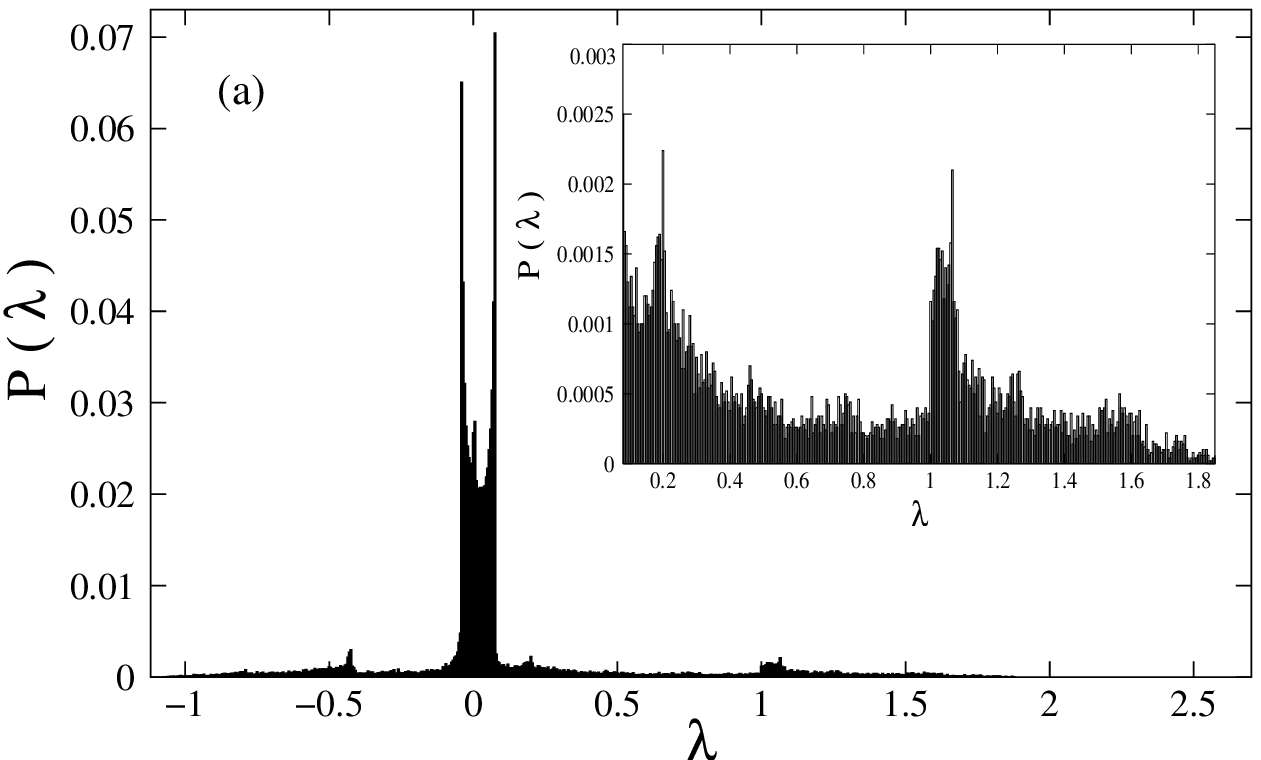} &
\includegraphics[height=6.5cm,width=8cm]{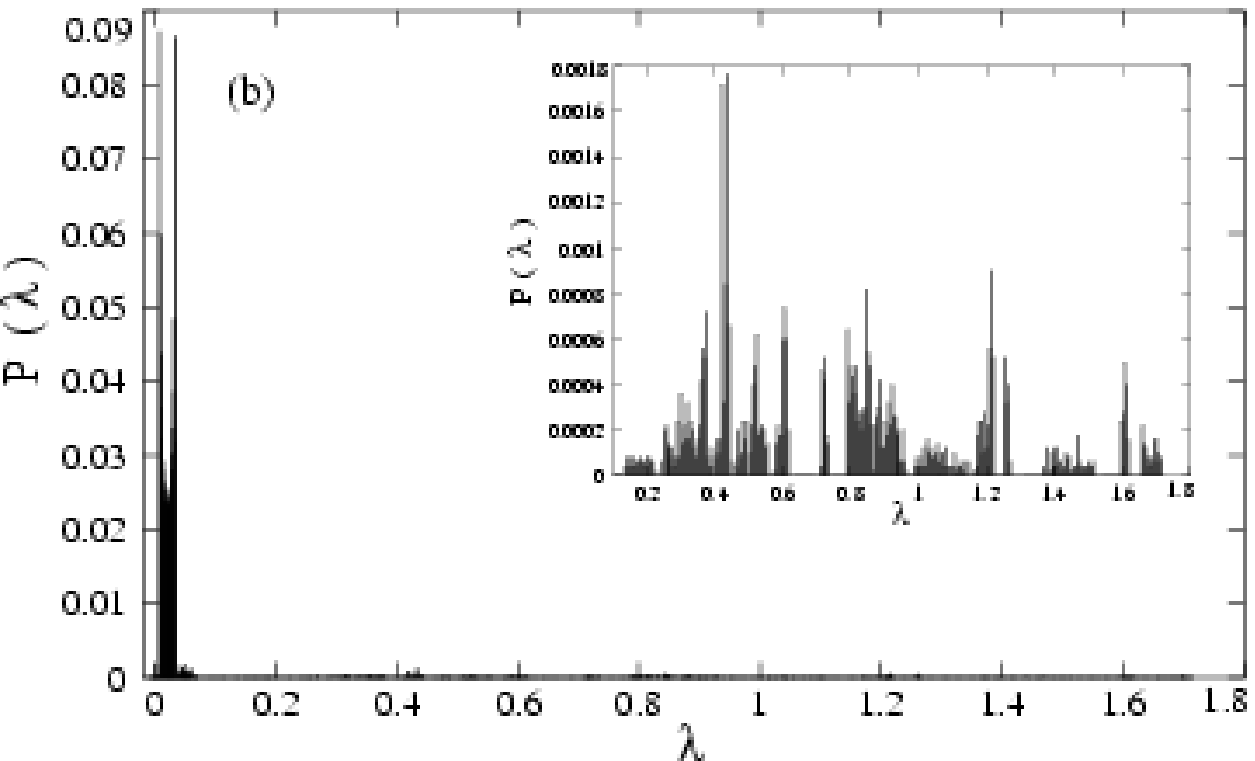}\\

\vspace{-.2in}
\end{tabular}            
\begin{tabular}{c}
\includegraphics[height=6.5cm,width=8cm]{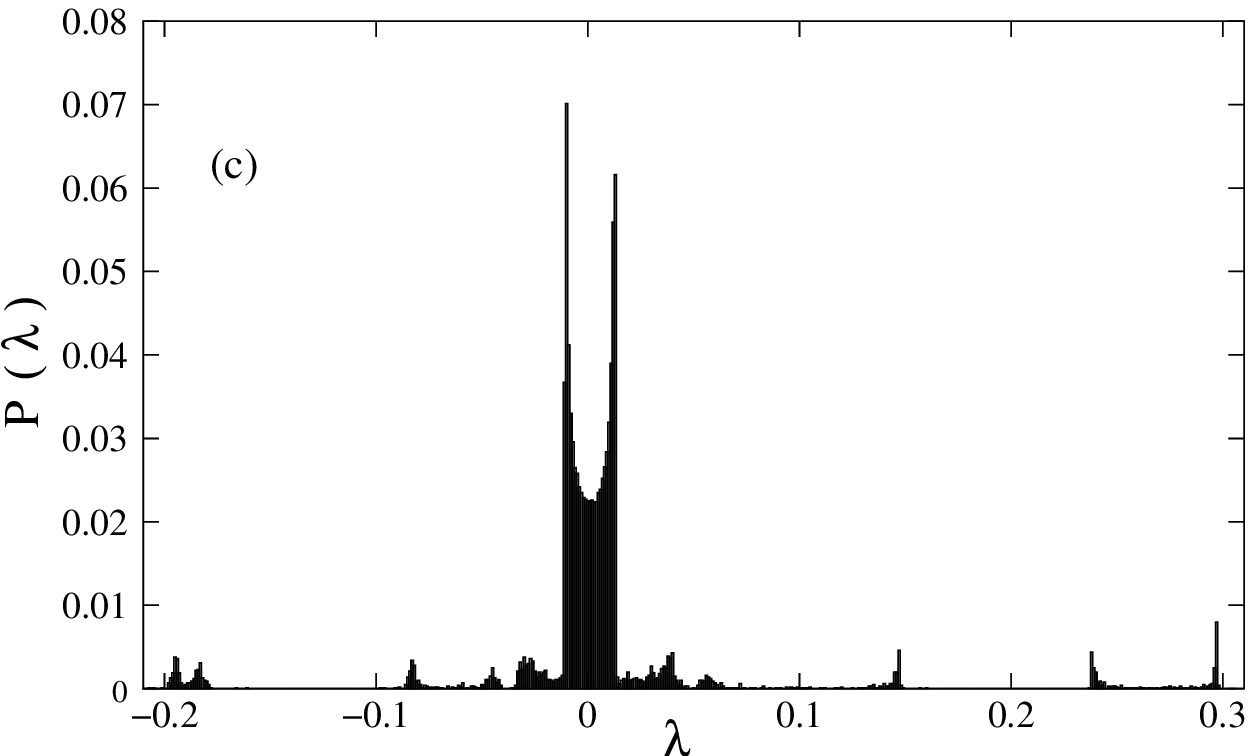}
\end{tabular}
\end{center}                                      
\caption{ shows the eigenvalue distribution (bin size=0.005) for (a) STI belonging to the DP class at $\Omega=0.06,\epsilon=0.7928$, (b) Spatial intermittency with quasi-periodic bursts at $\Omega=0.04,\epsilon=0.402$, and (c) SI with TW bursts at $\Omega=0.025, \epsilon=0.9496$. A section of the eigenvalue distribution is magnified in the inset figures. Gaps are seen in the spatial intermittency eigenvalue distributions whereas the eigenvalue distribution for STI does not show any such gaps.\label{nhgm}}
\end{figure}   


It has been seen earlier that the signature of the DP and non-DP behaviour in this system can be seen in the dynamical characterisers of the system, specifically in the eigenvalues of the one-step stability matrix. Regimes of STI with DP behaviour exhibited a continuous eigenvalue spectrum, whereas regimes of spatial intermittency showed an eigenvalue spectrum with level repulsion, where distinct gaps were seen in the spectrum.

The linear stability matrix of the evolution equation \ref{evol} at one time-step about the  solution of interest is given  by the  $N \times N$ dimensional matrix, $M_t^N$, given below 
\begin{displaymath}
\mathbf{M_t^N} = 
\left( \begin{array}{cccccc}
\epsilon_s f'(x_1^t) & \epsilon_n f'(x_2^t) & 0 & \ldots& 0 &\epsilon_n f'(x_N^t)\\
\epsilon_n f'(x_1^t)&\epsilon_s f'(x_2^t)&\epsilon_n f'(x_3^t)&0&\ldots&0\\
0&\epsilon_n f'(x_2^t)&\epsilon_s f'(x_3^t)&\ldots&0&0\\	
\vdots&\vdots&\vdots&\vdots&\vdots&\vdots\\
\epsilon_n f'(x_1^t)&0&\ldots&~~~0~~~&\epsilon_n f'(x_{N-1}^t)&\epsilon_s
f'(x_N^t)\\ 
\end{array}\right)
\end{displaymath}
 where, $\epsilon_s=1-\epsilon$, $\epsilon_n=\epsilon/2$, and $f'(x_i^t)=1-K\cos (2\pi x_i^t)$. $x_i^t$ is the state variable at site $i$ at time $t$, and a lattice of $N$ sites is considered. The diagonalisation of the stability matrix gives the $N$ eigenvalues at time $t$.

\subsection{The eigenvalue distribution}

\begin{figure}[!b]
\begin{center}
\begin{tabular}{cc}
\includegraphics[height=5.5cm,width=7cm]{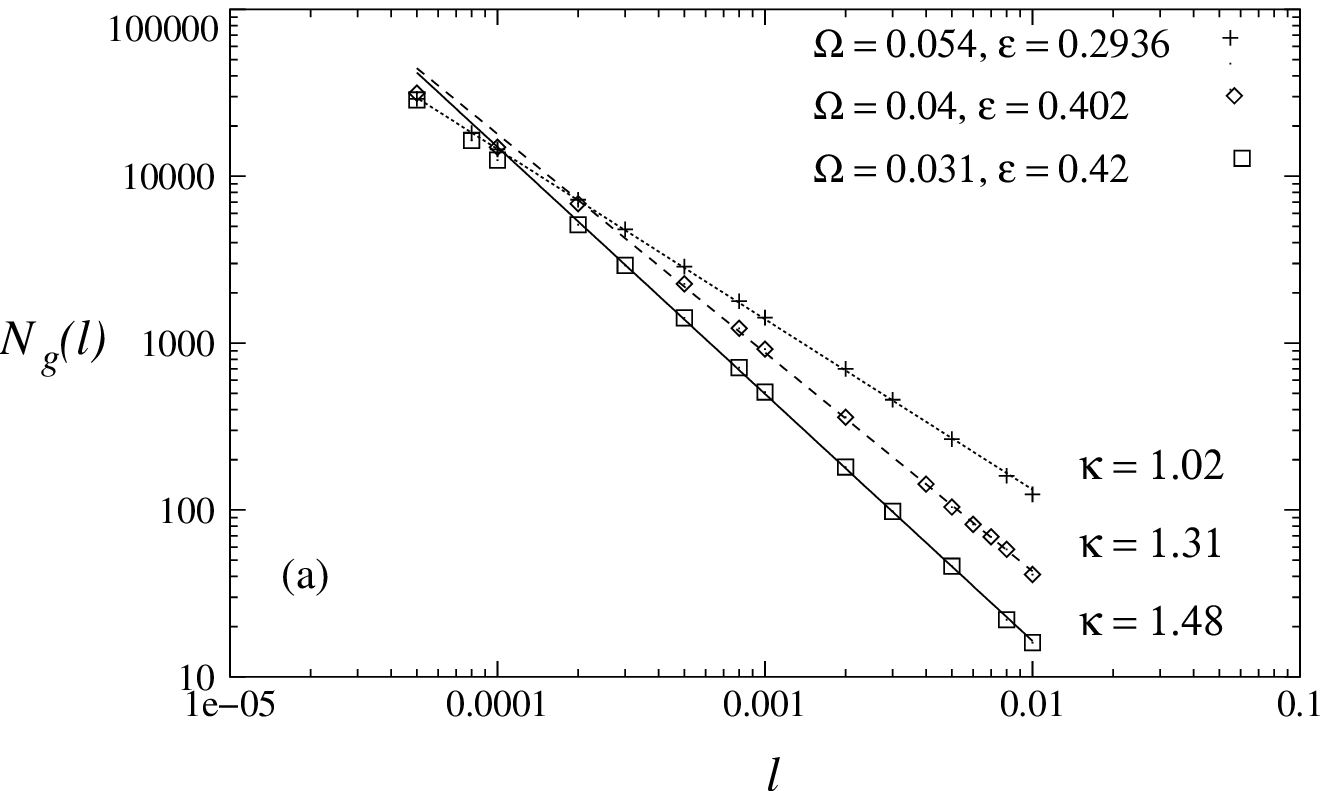} &\includegraphics[height=5.5cm,width=7cm]{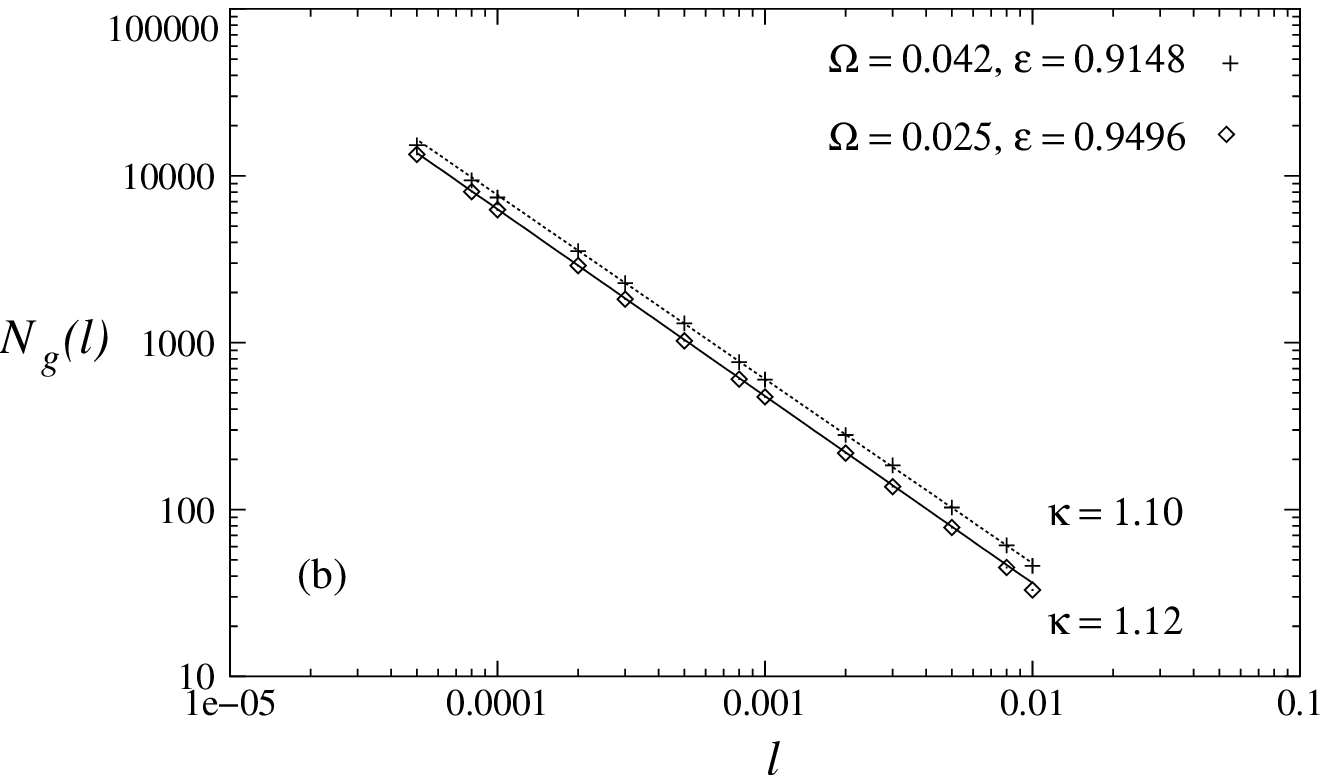}
\end{tabular}
\end{center}
\vspace{-.3in}
\caption{shows the number of vacant bins, $N_g(l)$ plotted against the bin size, $l$ on a log-log (base 10) scale for (a) SI with quasi-periodic bursts, and (b) SI with periodic bursts. The exponent, $\kappa$ associated with $N_g(l)\sim l^{-\kappa}$ is given in the figures. \label{ngl}}
\end{figure}

The eigenvalue distribution, $P(\lambda)$  of the matrix above, in all the 
cases studied here, have been obtained after averaging over 50 initial 
conditions for $1000$ lattice sites.
 Figure \ref{nhgm} shows 
the eigenvalue distributions for STI belonging to DP class at the typical parameter value $\Omega=0.06, \epsilon=0.7928$ (a),  SI with quasi-periodic bursts  at the typical value $\Omega=0.04, \epsilon=0.402$ (b), and SI with TW bursts at $\Omega=0.025, \epsilon=0.9496$ (c). The bin size chosen is 0.005. It is clear from the figure that the eigenvalue distribution of the DP class at this value of bin size is continuous whereas distinct gaps can be seen in the distribution for the spatial intermittency class for both quasi-periodic and periodic bursts.

For the case  of the SI, the number of vacant bins in the eigenvalue distribution, $N_g(l)$ scales as a power-law  $N_g(l)\sim l^{-\kappa}$ where $l$ is the bin-size (Figure \ref{ngl}). However, the exponent, $\kappa$ depends on the inherent dynamics of the burst states. The exponent, $\kappa$ for SI with quasi-periodic bursts have been listed in Table \ref{si1} and the exponents for SI with periodic bursts are listed in Table \ref{si2} .

Within the SI class, the value of $\kappa$ is seen to be stable within each period for the periodic bursts (See Fig. \ref{ngl}).  
In the quasi-periodic case, the natural frequencies of the dynamics are different 
at different values of the parameter and hence $\kappa$ values are different for different values of the parameters.  
It is also useful to track the temporal evolution of the largest eigenvalue to
identify the signatures of the differences between these three cases.
\subsection{The temporal evolution of the largest eigenvalue}
 \begin{figure}[!t]
\begin{center}
\begin{tabular}{cc}
\includegraphics[height=5.5cm,width=7.5cm]{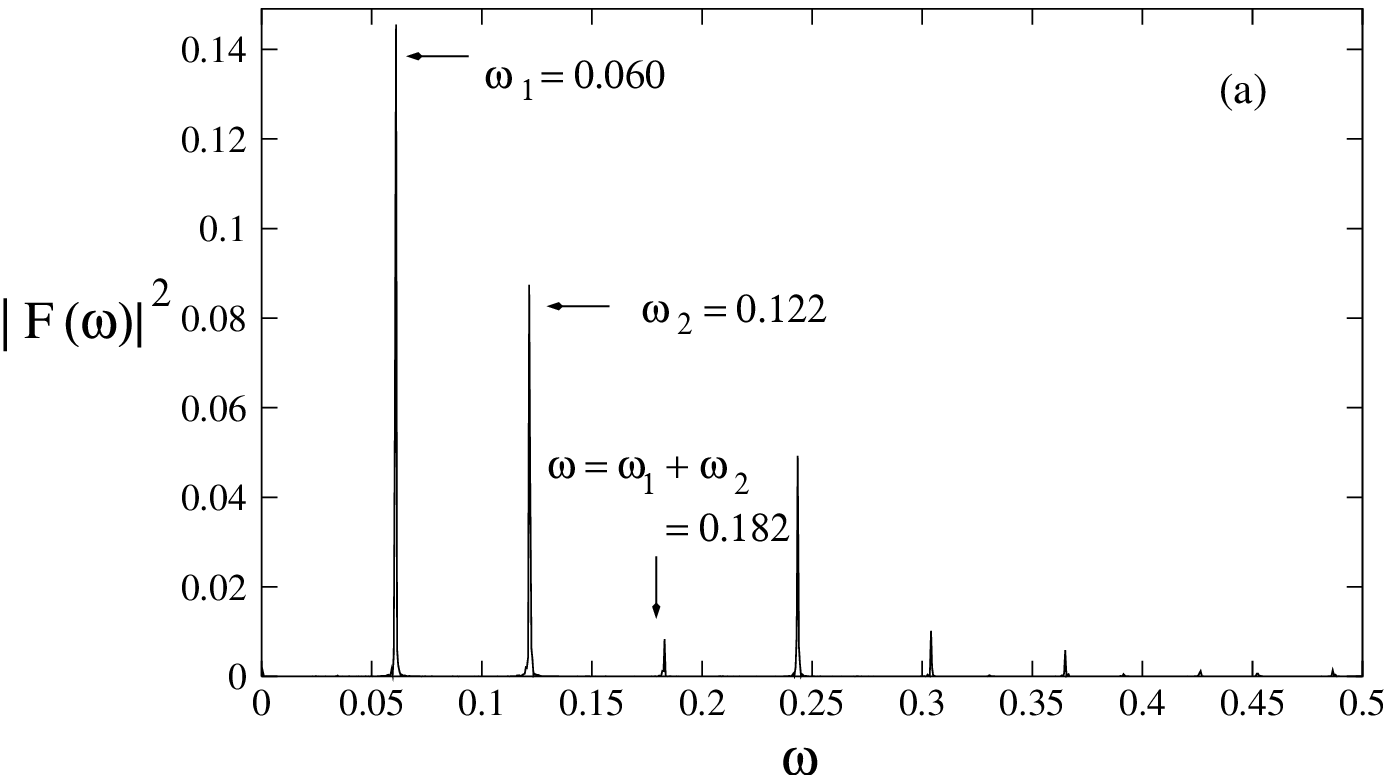} & \includegraphics[height=5.5cm,width=7.5cm]{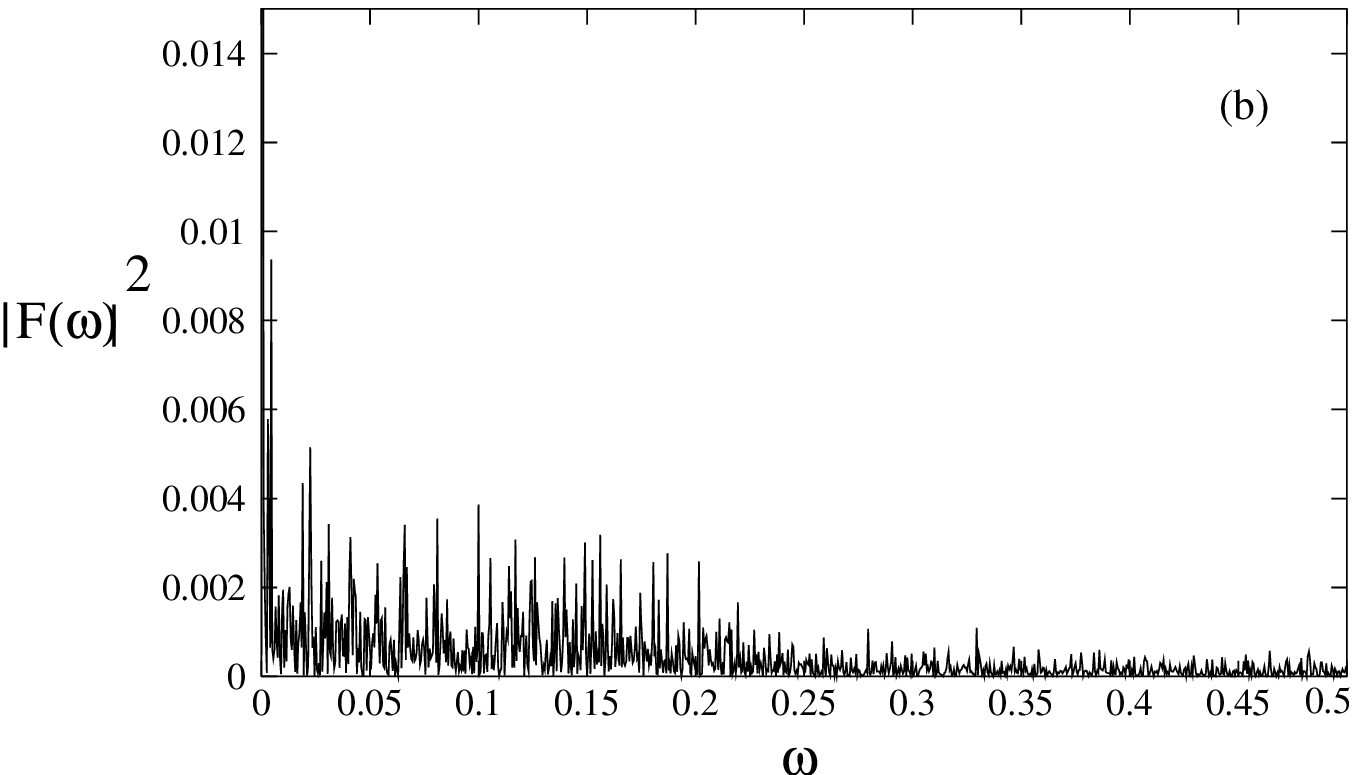}
\end{tabular}
\includegraphics[height=5.5cm,width=7.5cm]{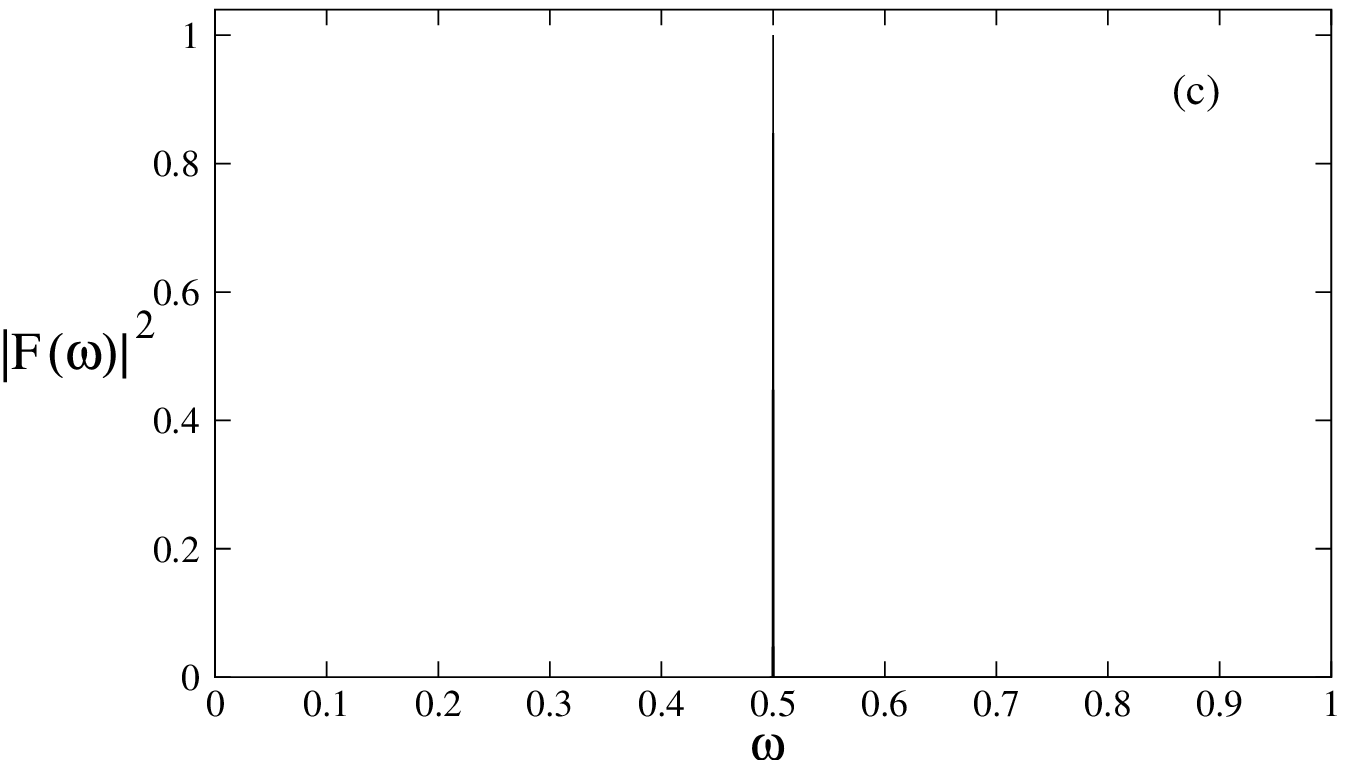}
\end{center}
\vspace{-.3in}
\caption{shows the power spectrum of the time series of the largest
eigenvalue, $\lambda_m(t)$ at (a) $\Omega=0.058,\epsilon=0.291$, where
SI with quasi-periodic behaviour is seen, (b)
$\Omega=0.06,\epsilon=0.7928$, where STI of the DP class is seen, and
(c) $\Omega=0.026, \epsilon=0.948$, where SI with TW bursts is seen.
Three main frequencies $\omega_1$, $\omega_2$ and $\omega_1+\omega_2$
are seen in (a) whereas, a broadband spectrum is seen in the case of STI
of the DP class as seen in (b). The spectrum of SI with TW shows a peak
at $0.5$ as seen in (c).\label{pslm}}
\end{figure}
The temporal evolution of the largest eigenvalue of the stability matrix,
$\lambda_{m}$ with $t$ contains information about the dynamical behaviour of the
burst states. The time series of the largest eigenvalue was obtained for the three
cases: STI of the DP class, SI with quasi-periodic bursts, and SI with TW bursts. After the initial transient, $\lambda_m$ settles down to the natural periods of the burst states. The power spectrum, $|F(\omega)|^2$ picks out the inherent frequencies in the system. This is evident from Figure \ref{pslm} in which the power spectrum of the time series of $\lambda_m(t)$ has been plotted as a function of the frequencies. 

In the case of SI with QP bursts (Figure \ref{pslm}(a)), peaks are seen at $\omega_1$, $\omega_2$, $\omega_1+\omega_2$ and at $m\omega_1+n\omega_2$ ($m, n > 0$) as is typical of a quasi-periodic behaviour. However, in the case of STI of the DP class , a broad band spectrum is obtained which implies that the burst states contains many frequencies. In the case of SI with periodic bursts, peaks are seen in the power spectrum at $\omega=0.5$ for SI with TW bursts (Figure \ref{pslm}(c)), indicating  period-2 temporal behaviour, whereas the peaks are seen at $\omega=0.2$, and $0.4$ for the type of SI in which the temporal behaviour of the bursts is period-5.

Thus we note that DP behaviour is associated with a broad-band spectrum for 
the power spectrum of the  temporal evolution of the largest eigenvalue, as well as a gapless distribution of eigenvalues, whereas the SI or non-DP behaviour is associated with the characteristic power spectrum of the temporal nature of the burst states i.e. periodic or quasi-periodic behaviour, and distinct gaps in the eigenvalue distribution.

\section{Conclusions}

Thus, spatiotemporal intermittency of several distinct types can be seen in different regions of the phase diagram of  the coupled sine circle map lattice.    
STI is seen all along the bifurcation boundaries of bifurcations from
the synchronized solutions. These bifurcations are of the
tangent-tangent (TT) and tangent-period doubling (TP) type. The
universal behaviour of the system as typified by the laminar length
exponents is of two types- the directed percolation (DP) class and the
non-DP class. STI with synchronized laminar states belongs convincingly
to the DP class and can be seen after both TT and TP bifurcations from
the synchronized state. This class of STI is remarkably free of the
solitons which spoil the DP behaviour in other models such as the
Chat\'e Manneville CML. Other regimes of the phase diagram show spatial
intermittency (SI) behaviour where the laminar regions show power-law
scaling and are periodic in behaviour, and the burst states show temporally 
regular behaviour of the periodic and quasi-periodic type. This type of 
intermittency is clearly not of the DP type and has earlier been seen in the 
inhomogenously coupled logistic map lattice.

 In addition to the  two regimes above, we also see STI with traveling wave (TW) laminar state in some regions of the parameter space. This kind of STI arises as the result of a TP bifurcation from SI with synchronized laminar state and TW bursts. This type of STI is contaminated with solitons and hence shows non-universal exponents. The soliton lifetimes depend on the parameter values and their distributions show two characteristic regimes. In the first regime, where typical lifetimes are short, the distribution peaks at short lifetimes showing the presence of a characteristic soliton lifetime scale but has a power-law tail. In the second regime, where soliton lifetimes are typically larger, the distribution has no characteristic scale and shows power-law behaviour with an exponent in the range $1.1-1.2$.

The dynamic characterisers of
 the system, namely, the eigenvalues of the stability matrix, shows signatures of these distinct types of behaviour. The DP regime is characterised by a gapless eigenvalue distribution and a broadband power spectrum of the time series of the largest eigenvalue. For the SI case, i.e. the non-DP regime, distinct gaps are seen in the eigenvalue distribution,  and the power spectrum of the temporal evolution of the largest eigenvalue is characteristic of periodic or quasi-periodic behaviour depending on the temporal nature of the burst states. 

The origin of the different types of universal behaviour in different parameter regimes appears to lie in the long range correlations in the system. These correlations in the system appear to change character in different regimes of parameter space leading to  dynamic behaviour with associated exponents of the DP and non-DP types. 
In order to gain insight into the nature of the
correlations in this system, and the way in which they change character in different parameter regimes, we plan to set up probabilistic cellular
automata which exhibit similar regimes and to examine their associated
spin Hamiltonians \cite{Domany}. 
Absorbing phase transitions are seen in other CML-s \cite{chatepmpm, chatepmpm1,chatejk} and in pair contact processes \cite{hinrich1,odor1}.
Models of non-equilibrium wetting set up using contact processes with long-range interactions also 
show DP or non-DP behaviour depending on the activation rate at sites at the edges of inactive islands \cite{ginelli}. Similar ideas may apply to the behaviour in our model as well. We hope to explore this direction in future work.

\section{Acknowledgments}

NG thanks DST, India for partial support. ZJ thanks CSIR, India for financial support.

\end{document}